\documentclass[aps, prl, twocolumn, superscriptaddress, amsmath, tightenlines, longbibliography]{revtex4-1}
\usepackage[none]{hyphenat}
\usepackage{amsmath}
\usepackage{amssymb}
\usepackage{amsfonts}
\usepackage{physics}
\usepackage{graphicx}
\usepackage{epsfig}
\usepackage{color}
\usepackage{textcomp}
\usepackage[colorlinks,citecolor=blue]{hyperref}
\usepackage{mathrsfs}
\usepackage{appendix}
\usepackage{dcolumn}
\usepackage{booktabs}
\usepackage{units}
\usepackage{url}
\usepackage[colorlinks]{hyperref}
\hypersetup{%
	plainpages=true,
	breaklinks=true,       %not default in dvips mode, so we must specify
	hypertexnames=false,  %not ideal, but needed when pagenums duplicate (`i' vs. `1')
	pageanchor=true,
	colorlinks=true,
	linkcolor={blue},
	citecolor={blue},
	urlcolor={blue},
	anchorcolor={black}
}
\begin{document}
\title{Collective Radiance of Giant Atoms in Non-Markovian Regime}
\author{Qing-Yang Qiu}
\affiliation{School of Physics, Huazhong University of Science and Technology, Wuhan, 430074, P. R. China}

\author{Ying Wu}
\affiliation{School of Physics, Huazhong University of Science and Technology, Wuhan, 430074, P. R. China}

\author{Xin-You L\"{u}}\email{xinyoulu@hust.edu.cn}
\affiliation{School of Physics, Huazhong University of Science and Technology, Wuhan, 430074, P. R. China}

\date{\today}% It is always \today, today,
\begin{abstract}
We investigate the non-Markovian dynamics of two giant artificial atoms interacting with a continuum of bosonic modes in a one-dimensional (1D) waveguide. Based on the diagrammatic method, we present the exact analytical solutions, which predict the rich phenomena of collective radiance. For the certain collective states, the decay rates are found to be far beyond that predicted in the the Dicke model and standard Markovian framework, which indicates the occurrence of super-superradiance. The superadiance-to-subradiance transition could be realized by adjusting the exchange symmetry of giant atoms. Moreover, there exist multiple bound states in continuum (BICs), with photons/phonons bouncing back and forth in the cavity-like geometries formed by the coupling points. The trapped photons/phonons in the BICs can also be re-released conveniently by changing the energy level splitting of giant atoms. The mechanism relies on the joint effects of the coherent time-delayed feedback and the interference between the coupling points of giant atoms. This work fundamentally broadens the fields of giant atom collective radiance by introducing non-Markovianity. It also paves the way for a clean analytical description of the nonlinear open quantum system with more complex retardation.
\end{abstract}
%corresponding excitations

\maketitle
Collective radiance\,\cite{M. Gross,Z. Wang,K. Sinha1} is a prototypical topic in the theoretical study of indirect interaction among emitters mediated by photons or phonons. Dicke revealed that the radiation emitted from an assembly of two-level atoms would be enhanced, and the total radiated power is proportional to square of the number of participating dipoles\,\cite{R. H. Dicke}. Generally speaking, the collective enhancement of radiative decay usually comes up when the emitters are very close to each other. However, there is still a case when the separation between neighboring emitters becomes comparable to the coherence length of an independently emitted photon. In this regime, the traveling time of a photon, normally ignored under the Markov approximation, is no longer negligible, since it will form a link among the emitters and then distinctly affect the interference properties of the electromagnetic field. To avoid the paradox caused by the Markovian approximation\,\cite{S. Longhi}, the exact theoretical description of non-Markov dynamics becomes very necessary.

It has been shown that the informational back-flow of the non-Markovian environment commonly originates from structured bath spectral densities\,\cite{N. Vats,A. Gonzlez-Tudela,A. Gonzlez-Tudela1,L.Xu} or strong system-bath couplings\,\cite{D. De Bernardis}. Non-Markovian retarded effects have been previously shown in scores of systems, including the geometrically large atomic system\,\cite{N. E. Rehler}, the single atom in front of a mirror\,\cite{T. Tufarelli}, the double giant cavities system\,\cite{Y.T. Zhu}, two macroscopically separated emitters\,\cite{K. Sinha}, and a linear chain of $N$ separated qubits\,\cite{F. Dinc}. In recent years, a rich variety of numerical and analytical methods, for instance, matrix product states, space-discretized waveguide\,\cite{S. A. Regidor,C. Guo,T. Lacroix,S. A. Regidor1}, and the diagrammatic method have been proposed to have an entire grasp of the counterintuitive phenomena induced by non-Markovianity\,\cite{F. Dinc1}.
\begin{figure*}[t]
  \centering
  \includegraphics[width=17cm]{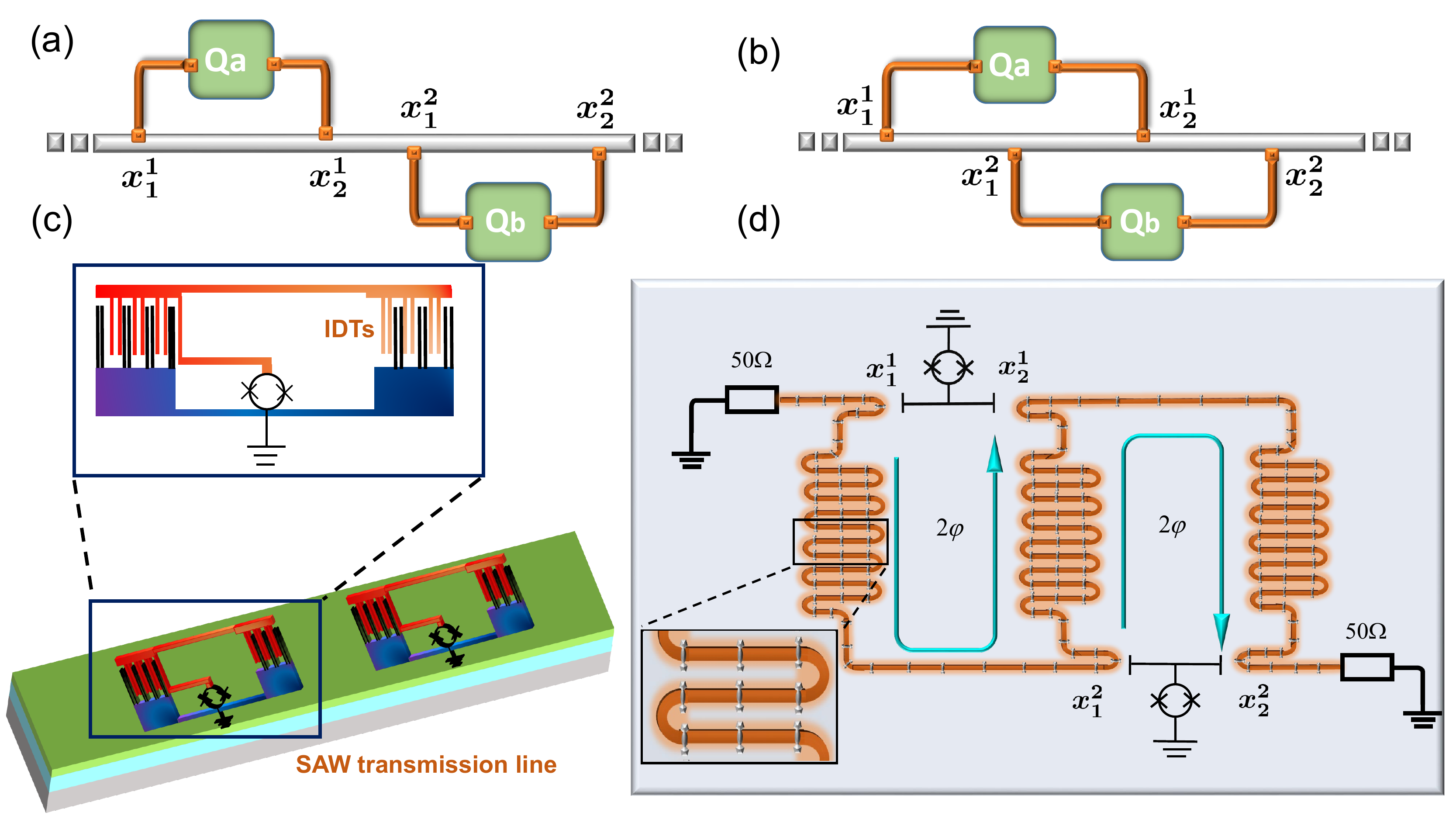}
  \caption{(Color online) (a, b) Two separate or braided superconducting qubits coupled to a 1D waveguide at four connecting points labeled by $x_{mn}$. The experimental implementations of the (c) separate and (d) braided configurations. (c) Two transmon qubits (superconducting quantum interference device sketched by black circle) are embedded in a surface acoustic waves transmission line with large interdigital transducer from it's two islands to generate and receive acoustic signals. (d) Two X-mon qubits coupled capacitively to a meandering microwave coplanar waveguide.}\label{fig1}
\end{figure*}
In conventional formalism of light-matter interaction, atoms are generally treated as point-like dipoles. Recently, following the tremendous progress in experimental physics for superconducting circuits\,\cite{X. Gu,J. Q. You,P. Krantz,P.Zhao}, atomic size in the platforms with surface acoustic waves\,\cite{M. V. Gustafsson,G. Andersson,L. Guo1} or meandering coplanar waveguide\,\cite{B.Kannan} can be comparable to the wavelength of microwave photons they interact with, and thus invalidates the dipole approximation. Artificial atoms with such a novel structure are called superconducting giant atoms. The nature of non-local coupling of giant atoms has led to many interesting effects, including frequency-dependent decays and Lamb shifts\,\cite{A. F. Kockum}, single-photon scattering\,\cite{W. Zhao,X.-L. Yin}, tunable chiral bound states\,\cite{X. Wang}, oscillating bound states\,\cite{L. Guo}, unconventional electromagnetically induced transparency\,\cite{Y.T. Zhu1}, and decoherence-free atomic interaction\,\cite{A. F. Kockum1}. While the non-Markovian dynamics of a single two-legged giant atom can be well described analytically through the Lambert W function\,\cite{F. M. Asl}, the collective performance of many giant atoms with retarded feedback is yet perplexing.

Here we study the collective radiance of two giant artificial atoms in the 1D waveguide quantum electrodynamics (QED) system\,\cite{Q. Li,L. Zhou,X.-L. Dong,DD,ll}. We find the analytical solutions of the atomic excitation probability amplitudes by solving the relevant multi-delay differential equations with the diagrammatic method. For the purpose of further insights of collective behavior in the non-Markovian regime, we also present the exact atomic collective decay rates of the considered model, where super-superadiance\,\cite{F. Dinc2} can be observed for certain critical atomic separation. The complex interference effects of delayed photons induce a wealth of dynamical behaviors, including the multiple enhanced radiation bursts and long-lived BICs. Interestingly, those collective radiance dynamics, e.g., the superadiance-to-subradiance transition and the change of BIC areas, are tunable with respect to the initial states, the distance between the coupling points and the atomic geometric structures. Moreover, we also show that the trapped photons/phonons within BIC can be re-released by applying a time-dependent energy level splitting on the giant atoms.

\emph{Model and multi-delay differential equations.}---In Fig.\,\ref{fig1}, we show two types of theoretical models and the corresponding experimental implementations for the giant atoms. By neglecting counter-rotating terms and applying electric-dipole approximation, the total Hamiltonian of the system  reads ($\hbar=1$)
\begin{align}
&\!\!\!\hat{H}=\frac{\omega_{0}}{2}\underset{m=a,b}{\sum}\hat{\sigma}_{+}^{(m)}\hat{\sigma}_{-}^{(m)}
\!+\!\underset{\alpha=R,L}{\sum}\int_{0}^{+\infty}\frac{\omega}{2}\hat{a}_{\alpha}^{\dagger}(\omega)\hat{a}_{\alpha}(\omega)d\omega\nonumber\\
&\!\!\!+\underset{\alpha=R,L}{\sum}\sum_{m=a,b}\sum_{n=1,2}\int_{0}^{+\infty}\!g(\omega)\hat{\sigma}_{+}^{(m)}\hat{a}_{\alpha}(\omega)e^{i\epsilon_{\alpha}\omega x_{mn}/v_{g}}d\omega+\text{H.c.},\label{eq1}
\end{align}
where $\hat{a}_{\alpha}(\omega)$ is the annihilation operator of the right- $(\alpha=R)$ or left- $(\alpha=L)$ moving photons (or phonons) satisfying $[\hat{a}_{\alpha}(\omega),\hat{a}^{\dagger}_{\alpha'}(\omega')]=\delta_{\alpha\alpha'}\delta(\omega-\omega')$. $x_{mn}$ denotes the $n$th coupling point of $m$th giant atom, and $v_{g}$ is the group velocity of the field in the waveguide. The first and second terms on the right hand represent the free Hamiltonians of giant atoms and 1D bosonic modes, respectively. The resonant frequency of emitters $\omega_{0}$ is assumed to be far away from the cut-off frequency of the waveguide, which confirms that a perfect linearized dispersion can be applied. Moreover, $\hat{\sigma}_{+}^{(m)}=|e\rangle_{mm}\langle g|$ is the raising operator of $m$th emitter, and the notations $\epsilon_{L}=-1$ and $\epsilon_{R}=+1$ have been introduced in the third term.

To obtain the non-Markov dynamical property of the system, we proceed by analytically determining the excitation probability amplitudes of two giant atoms. The following discussion have been limited into a single-excitation subspace, and then the instantaneous state is
\begin{align}\label{eq2}
\bigl|\psi(t)\bigr\rangle=&\sum_{m=a,b}c_{m}(t)\hat{\sigma}_{+}^{(m)}\bigl|g,g,\{0\}\bigr\rangle\nonumber\\
&+\underset{\alpha=R,L}{\sum}\int_{0}^{+\infty}d\omega
\varphi_{\alpha}(\omega,t)\hat{a}_{\alpha}^{\dagger}(\omega)\bigl|g,g,\{0\}\bigr\rangle,
\end{align}
where $\bigl|g,g,\{0\}\bigr\rangle$ is the ground state of system and $\bigl|\{0\}\bigr\rangle$ denotes the vacuum state of 1D waveguide modes. Moreover, $\varphi_{L/R}(\omega,t)$ represents the amplitude of probability to find a single photon propagating to left/right at time $t$ with mode $\omega$ in the waveguide. In the interaction picture, the system dynamics is decided by the schr$\rm{\ddot{o}}$dinger equation $i\bigl|\dot{\psi}(t)\bigr\rangle=\hat{H}_{\rm{int}}\bigl|\psi(t)\bigr\rangle$, where $\hat{H}_{\rm{int}}$ is the interaction Hamiltonian of the system. Tracing out the bosonic modes, then we obtain the equations of motion (EOMs) of separate giant atoms, and it can be described as (see Supporting Information S1 for more details)
\begin{align}\label{eq3}
\dot{c}_{m}(t)=-\gamma c_{m}(t)-F_{m,1}(t)-F_{m,2}(t)-\frac{1}{2}[F_{n,1}(t)+F_{n,3}(t)]
\end{align}
for $m\neq n$. Similarly, the atomic evolution of excitation amplitudes of braided giant atoms is governed by\,\cite{LL}
\begin{align}\label{eq4}
\!\!\!\dot{c}_{m}(t)=&-\gamma c_{m}(t)-F_{m,2}(t)-\frac{1}{2}[3F_{n,1}(t)+F_{n,3}(t)],\!\!
\end{align}
\begin{figure}
  \centering
  \includegraphics[width=8.5cm]{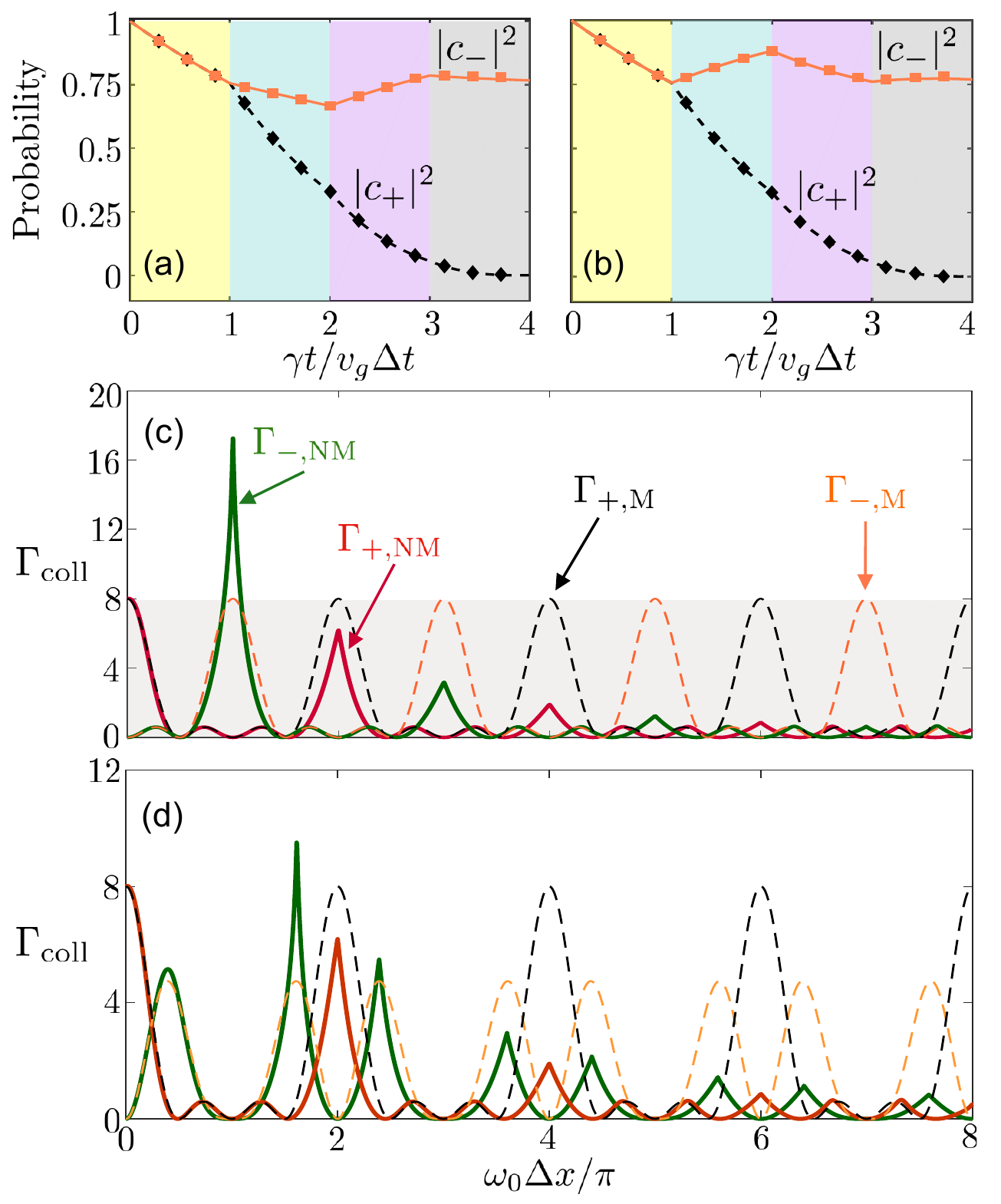}
  \caption{(Color online) The atomic excitation population $|c(t)|^{2}$ of two (a) separate and (b) braided giant atoms initially prepared in symmetric state (black diamonds and lines) and antisymmetric state (orange squares and lines). The analytical and numerical solutions are shown by the diamonds/squares and lines, respectively. The corresponding parameters are $\varphi=2n\pi$($\varphi=(2n+1)\pi$) for state $|+\rangle$($|-\rangle$)in (a) and $\varphi=2n\pi$($\varphi=2n\pi$) for state $|+\rangle$($|-\rangle$)in (b). Other parameter used here is $\eta=\gamma\Delta t=0.15$. The Markovian (dashed line) and non-Markovian (solid line) collective decay rates of braided (c) and separate (d) giant atoms versus $\omega_{0}\Delta x/\pi$ where $\omega_{0}=50\gamma$. Where the notation $\Gamma_{+/-,\rm{M/NM}}$ denotes the collective decay rate of symmetric/anti-symmetric state in the Markovian /non-Markovian regime. The shading region indicates the regime of $\Gamma_{\rm coll}<7.93$, and $\Gamma_{\rm coll}=7.93$ is the maximal non-Markovian decay rate for $4$ small atoms. Note that the phase shifts $\varphi$ are changed by adjusting the atomic resonance frequency $\omega_{0}$ in (a) and (b), while the ones in (c) and (d) are varied through the distance $\Delta x$.}\label{fig2}
\end{figure}
\noindent where $F_{m,n}(t)\equiv\gamma e^{in\varphi}c_{m}(t-n\Delta t)\Theta(t-n\Delta t)$ and $\varphi\equiv\omega_{0}\Delta t =k_{0}\Delta x$ is the field propagation phase difference between two adjacent legs. Here, the Heaviside step function $\Theta(\bullet)$ is introduced to hold causality and describes the time-delayed feedback of information among the multiple legs. The equidistant distribution of coupling points and the homogeneous decay rate $\gamma$ in each connecting point have been considered for simplicity. We also have assumed a flat spectral density of the waveguide modes around the resonance of the giant atoms, i.e., $g(\omega)\approx g(\omega_{0})=\sqrt{\gamma/(4\pi)}$.

The first term on the right-hand side of Eq.~(\ref{eq4}) depicts the spontaneous emission processes due to the Markovian dynamics. The second term stems from the nature of nonlocal coupling of a single giant atom, and the remainder describes the atomic relaxation dynamics controlled by the delayed photons released from another emitter. A differential equation with a series of information feedbacks like Eqs.(\ref{eq3})-(\ref{eq4}) is mathematically called a multi-delays differential equation. Here, we obtain the exact analytical description of the system's non-Markovian dynamical evolution via a diagrammatic method\,\cite{F. Dinc1}, which is excellently consistent with the fully numerical simulation. After some algebra, the general expression for the probability amplitude of two-atom-state is given by (see Supporting Information S2 for more details)
\begin{align}\label{eq5}
&c(t)=\frac{1}{2}\sum^{\infty}_{l=0}K_{l}(t)\sum_{s=0}^{l}D_{ls}(t)f_{s},
\end{align}
where $D_{ls}(t)\!\equiv\!\gamma^{l}(t-s\Delta t)^{l}$ and $K_{l}(t)\!\equiv\!\sqrt{2}c_{a}(0)\times$\, $e^{-(\gamma+i\omega_{0})(t-l\Delta t)}\Theta(t-l\Delta t)$ are defined for conciseness, and $f_s$ closely relies on both the specific configurations of the giant atoms and the atomic initial states. When the giant atoms are initially in the symmetric and antisymmetric states, i.e., $|\psi(0)\rangle=\ket{\pm}=\frac{1}{\sqrt{2}}(\left|eg\right\rangle \pm\left|ge\right\rangle )$, the corresponding amplitude $c_{\pm}(t)$ can be given by Eq.\,(\ref{eq5}). In Figs.\,\ref{fig2}(a) and (b), we plot the atomic amplitude $|c_{\pm}(t)|^2$ in the early four basic traveling time ($4\Delta t$), which clearly shows the consistent analytical (dotted line) and numerical (solid line) results. Then the above analytical expression can be used to simulate other interesting phenomena induced by coherent time-delayed feedback in the later discussion.

\emph{Collective decay rates.}---We next study the collective decay rates of two giant atoms that interact with a common radiation field. The interference between emissions from the coupling points of the giant atoms may enhance or suppress the decay for certain states. This phenomenon is famous as super/subradiance\,\cite{R. H. Dicke}. Here, we utilize the real-space approach to describe quantitatively the collective decay rates. The system Hamiltonian (\ref{eq1}) in the real-space becomes ($v_{g}=1$)
\begin{align}\label{eq6}
\!\!\hat{H}\!=\!&\sum_{m=a,b}\omega_{0}\hat{\sigma}_{+}^{(m)}\hat{\sigma}_{-}^{(m)}\!+\!i\!\int \! dx\left[\hat{c}_{L}^{\dagger}(x)\frac{\partial}{\partial x}\hat{c}_{L}(x)\!-\!\hat{c}_{R}^{\dagger}(x)\frac{\partial}{\partial x}\hat{c}_{R}(x)\right]\nonumber\\
&\!+\!\sum_{m=a,b}\sum_{n=1,2}\!\sqrt{\frac{\gamma}{2}}\!\int \! dx\delta(x-x_{mn})\!\left[\!\sum_{\alpha}\!\hat{c}_{\alpha}^{\dagger}(x)\hat{\sigma}_{-}^{(m)}\!+\!\rm{H.c.}\right]\!,\!\!\!
\end{align}
where $\hat{c}_{\alpha}(x)$ is the annihilation operator for left- ($\alpha=L$) or right- ($\alpha=R$) moving photons at position $x$. The light-matter interactions take the form of delta functions, corresponding to the barriers located at the coupling points. Based on the similar strategy in\,\cite{F. Dinc2,H. Zheng}, the complete set of collective decay rates of arbitrarily separated giant atoms is given by the poles of the scattering parameters, which are governed by a transcendental equation (see Supporting Information S3 for
\begin{figure}
  \centering
  \includegraphics[width=8.8cm]{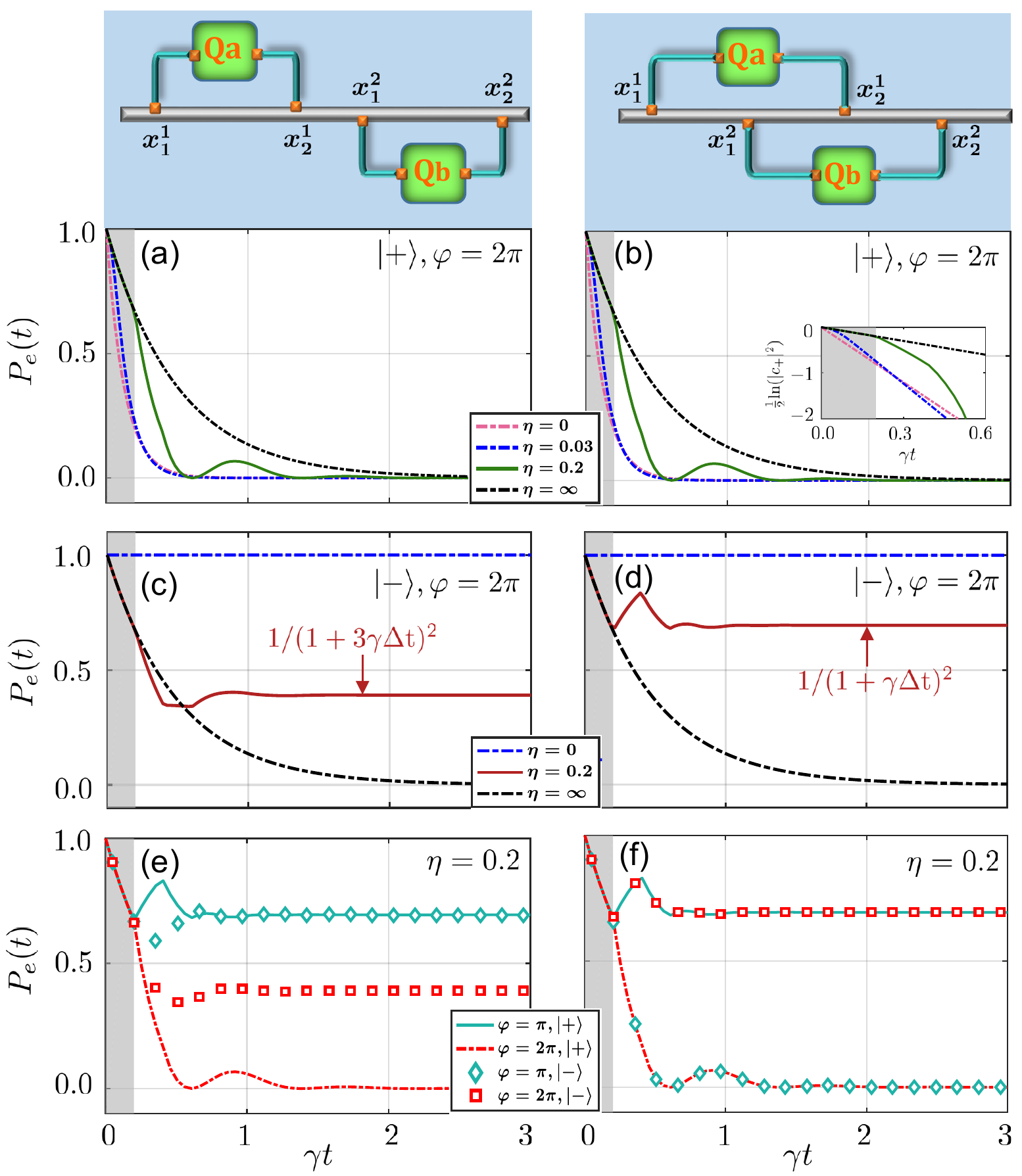}
  \caption{(Color online) Time evolution of atom excitation probability for the separate (left column) and braided (right column) configuration. (a,b) and (c,d) show respectively the evolution of symmetric and anti-symmetric state denoted by $|+\rangle$ and $|-\rangle$ with fixed light propagation phase $\varphi$ and different values of retardation $\eta$. The inset figure in (b) show the corresponding instantaneous collective decay rates. (e, f) depict how the parameters $\varphi$ and initial states influence the atomic evolution under a finite time delay $\eta$. The area filled by grey denotes a process in which photons emitted from atomic ensemble independently with $\eta=0.2$.}\label{fig3}
\end{figure}
\noindent  more details). Such a nonlinear characteristic equation can be reduced as a polynomial equation in the presence of the negligible delay, which gives the analytical solutions of the Markovian collective decay rates. More concretely, $\Gamma_{+,{\rm{M}}}=(2+3e^{i\varphi}+2e^{2i\varphi}+e^{3i\varphi})\gamma,\,\Gamma_{-,{\rm{M}}}=(2+e^{i\varphi}-2e^{2i\varphi}-e^{3i\varphi})\gamma$ for separate giant atoms, and
$\Gamma_{+,{\rm{M}}}=(2+3e^{i\varphi}+2e^{2i\varphi}+e^{3i\varphi})\gamma,\,\Gamma_{-,{\rm{M}}}=(2+2e^{2i\varphi}-3e^{i\varphi}-e^{3i\varphi})\gamma$ for braided giant atoms, where $\Gamma_{+/-,{\rm{M}}}$ denotes the collective decay rate of the symmetric/anti-symmetric state in the Markovian regime.

We plot the analytical Markovian collective decay rates $\Gamma_{\pm,\rm{M}}$ and the numerical non-Markovian collective decay rates $\Gamma_{\pm,\rm{NM}}$ for both braided and separate giant atoms in Fig.\,\ref{fig2}(c)-(d) by gradually increasing $\Delta x$. Firstly, it is shown that the Markovian collective decay rates guide effectively the real behavior of collective dynamics if the connecting points are slightly separated. When the distance $\Delta x$ is increased to a critical value $\Delta x_{c}$, the superradiant collective decay rate extends far beyond than that revealed by the Markovian approximation. Secondly, in the large separation limit, both the symmetric and antisymmetric non-Markovian collective decay rates drastically deviate from the Markovian ones, and tend to be subradiant. This suppression of decay can be explained as follows. After a giant atom finishes the initial emission from its coupling points, it can be re-excited by the wavepackets reflected from all other coupling points. The process of looping emission-reexcitation is repeatedly cycled over and over again and then the light needs longer time to struggle to free itself, which prolongs the effective lifetime of the emitters\,\cite{H. Zheng}. Lastly, the $2N^{2}\gamma$ ($N$ is the total connecting legs of a single giant atom) scaling of the maximum Markovian decay rate for two giant atoms surpasses the scaling of $4\gamma$ given by well-known Dicke superradiance for four small atoms. More interestingly, this scaling is further improved by introducing self-consistent coherent time-delayed feedback. Here we obtain that the super-superradiant decay rate reads about $9.51\gamma$ for separate atoms and $17.26\gamma$ for braided atoms. Different from the case of traditional ``small" atoms, the condition $\omega_{0}\Delta x_{c}=n\pi$ is not necessary to achieve a maximum superradiance.\vspace{0.15cm}

\emph{Collective emission dynamics and BICs.}---For illustration purposes, we further show the time evolution of the atomic excitation probability in Fig.\,\ref{fig3}. Collective emission dynamics can be significantly modified by the geometric structures of the giant atoms, retardation $\eta\equiv\gamma\Delta t$, atomic initial states, and the phase $\varphi$ of the electromagnetic field acquired upon propagation. In Figs.\,\ref{fig3}(a) and (b), we show the dynamics of symmetric state $|+\rangle$ with $\varphi=2n\pi$. By adjusting the distance of adjacent coupling legs from zero ($\eta=0$) to a slightly separated length (e.g., $\eta=0.03$), atomic evolution will go through from an exponential behavior ($\Gamma$=$8\gamma$) to a non-exponential behavior, where a given average decay rate may be drastically beyond $8\gamma$. As the retardation gets a further increase (e.g., $\eta=0.2$), oscillating behaviour of the atomic excitation amplitudes can be captured as a remarkable symbol of the non-Markovian recovery phenomenon. Finally, the atomic time evolution tends to become independent radiation ($\Gamma$=$2\gamma$) with a large atomic separation ($\eta=\infty$). Besides, one may note that both the separate and braided giant atoms share a same dynamical process if the atomic initial state is $|+\rangle$.

Figs.\,\ref{fig3}(c) and (d) show the subradiant emission dynamics of anti-symmetric state $|-\rangle$ with $\varphi=2n\pi$. In this case, the emitters can be frozen in the excited states as the retardation is set to be zero. For a finite delay (e.g., $\eta=0.2$), the atomic excitation probability may rest on a fixed value in the long time limit and maintain dynamic equilibrium with surrounding trapped photons. Specifically, this probability reaches the asymptotic value $(1+3\eta)^{-2}$ and $(1+\eta)^{-2}$ for separate giant
\begin{figure}
  \centering
  \includegraphics[width=8.5cm]{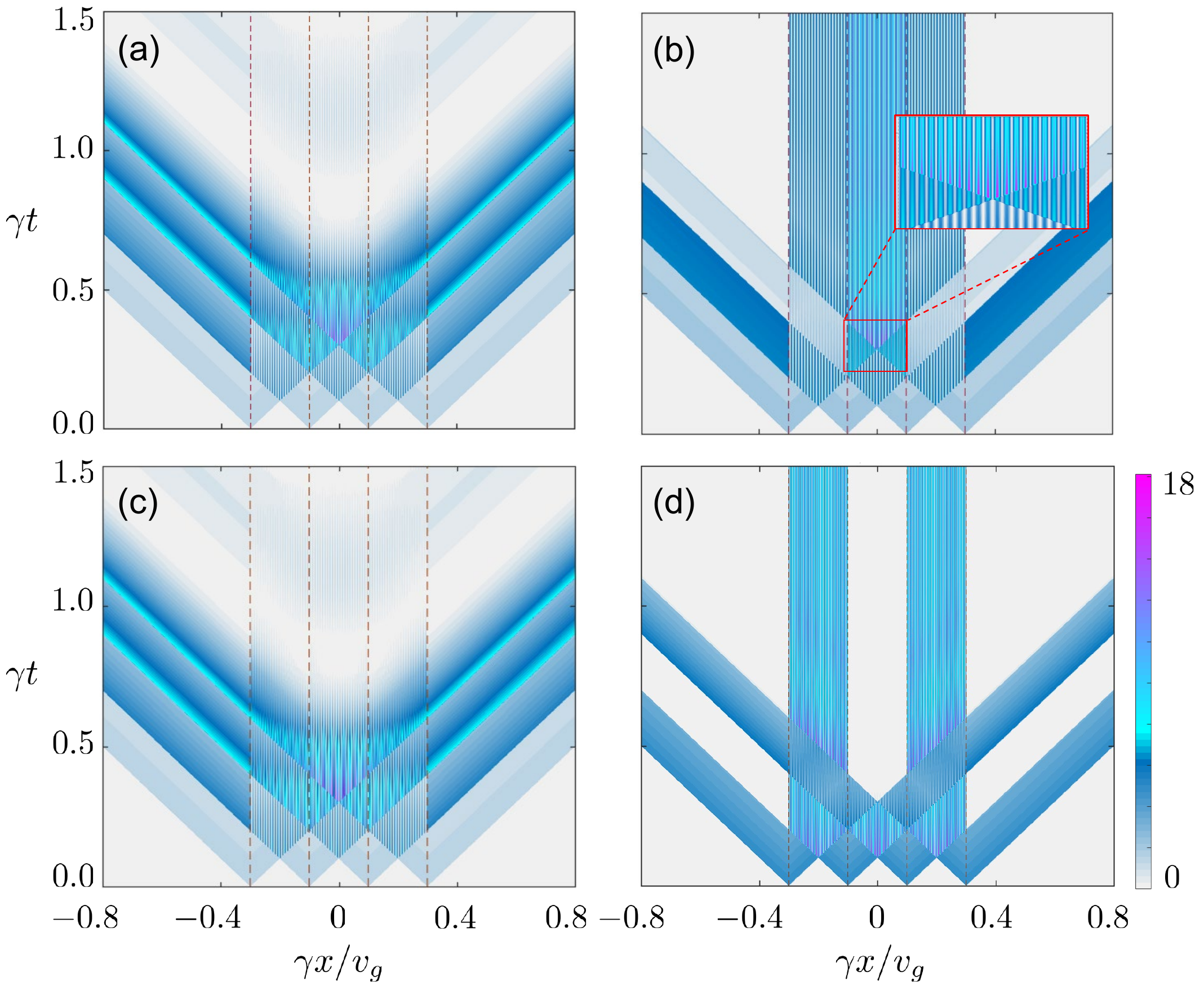}
  \caption{(Color online) Bosonic field density distribution of the system with two separate (a,b) or braided (c,d) giant atoms coupled to a 1D waveguide at four positions marked by brown dotted lines, i.e., $(x_{a1},x_{a2},x_{b1},x_{b2})=(-0.3,-0.1,0.1,0.3)v_{g}/\gamma$ for (a,b) and $(-0.3,0.1,-0.1,0.3)v_{g}/\gamma$ for (c,d). In our simulations, the atomic initial states are prepared to be exchange symmetric (a,c) and antisymmetric (b,d). Other parameters are $\varphi=2\pi,\eta=0.2,v_{g}=1$.
  }\label{fig4}
\end{figure}
\noindent atoms and braided giant atoms, respectively. In Figs.\,\ref{fig3}(e) and (f), we demonstrate that how the atomic evolution is manipulated by both $\varphi$ and initial states for a given retardation. In the case of separate giant atoms, the symmetric state $|+\rangle$ will transfer from a non-radiative state to a radiative state, and the anti-symmetric state $|-\rangle$ is going to jump to a lower populated steady state, while the phase shifts from $2(n+1)\pi$ to $2n\pi$. In another case of braided giant atoms, perfect synchronization in the time domain occurs by exchanging both the initial state ($|+\rangle\rightarrow |-\rangle$) and the phase ($n\pi\rightarrow 2n\pi$).

Moreover, it also can be shown from Fig.\,\ref{fig3} that, for a finite retardation $\eta>0$, a purely exponential decay can be observed before $t=\Delta t$ during which no delayed signals are received. After this independent emission process, one may discover multi-level collective radiation enhancement as soon as $t=n\Delta t$ due to the constructive multi-path interference between the spontaneous radiation and stimulated radiation. Or conversely, the system would evolve spontaneously into a deterministic $\rm{BIC}$ if $\langle \rm{BIC}|\psi(t\!=\!0)\rangle\!\neq\! 0$. Long-lived bound states in the continuum are waves that remain localized even though they coexist with an extended spectrum of radiating waves that can carry excitations away\,\cite{Hsu,F. H. Stillinger,D. C. Marinica,M. I. Molina}. Two conventional approaches to populate these states in this regard are spontaneous dissipation of initially excited atoms\,\cite{K. Sinha,L. Guo,T. Tufarelli} and multi-photon scattering on an unexcited atomic system\,\cite{G. Calaj,J.-T. Shen,R. Trivedi}. Here, we adopt the former way to generate system BICs during which the giant atom legs behave equivalently as effective mirrors. Then the giant atoms and surrounding trapped photons ultimately reach the dynamic equilibrium. Physically, such bound states stem from the destructive interference between the spontaneous radiation, reabsorption and stimulated radiation as described in Eqs.\,(\ref{eq3})-(\ref{eq4}). When the giant atoms are initially prepared in an anti-symmetric state, the analytical descriptions of BICs as shown in Supporting Information S4 are given by
\begin{align}
\!\!\!&|\psi\rangle^{\rm{b}}_{\rm{BIC}}\!=\!\frac{1}{\sqrt{1+\gamma\Delta t}}\{\frac{1}{\sqrt{2}}(|eg\rangle\!-\!|ge\rangle)\otimes|\rm{vac}\rangle\nonumber\\
&\!-\!\sqrt{\frac{\gamma v_{g}}{2\pi}}\int_{0}^{\infty}\!dk\frac{\sin \frac{3}{2}kd\!-\!\sin \frac{1}{2}kd}{\omega-\omega_{0}}\![\hat{a}^{\dagger}_{R}(\omega)\!-\!\hat{a}^{\dagger}_{L}(\omega)]|\emptyset\rangle\}\label{eq7}\\
\!\!\!&|\psi\rangle^{\rm{s}}_{\rm{BIC}}\!=\!\frac{1}{\sqrt{1+3\gamma\Delta t}}\{\frac{1}{\sqrt{2}}(|eg\rangle\!-\!|ge\rangle)\otimes|\rm{vac}\rangle\nonumber\\
&\!-\!\sqrt{\frac{\gamma v_{g}}{2\pi}}\int_{0}^{\infty}\!dk\frac{\sin \frac{3}{2}kd\!+\!\sin \frac{1}{2}kd}{\omega-\omega_{0}}[\hat{a}^{\dagger}_{R}(\omega)\!-\!\hat{a}^{\dagger}_{L}(\omega)]|\emptyset\rangle\},\!\label{eq8}
\end{align}
where $|\psi\rangle^{\rm{s/b}}_{\rm{BIC}}$ denotes the BIC in a system formed by separate/braided giant atoms coupled to a 1D waveguide, $d\equiv \Delta x$, and $|\emptyset\rangle$ is the ground state of the total system. As shown in Eqs.\,(\ref{eq7}) and (\ref{eq8}), we found quantitatively a significant overlap between BIC and the initial states in the long time limit. Specifically, we obtain that $c_{-}(t\rightarrow\infty)=|\langle\psi_{\rm{BIC}}|-\rangle|^{2}$ with a value $1/(1+3\gamma\Delta t)$ for separate atoms and $1/(1+\gamma\Delta t)$ for braided atoms, which are consistent with the calculations given by the final value theorem in the Laplace space (see Supporting Information S1 for more details).\vspace{0.15cm}

\emph{Multi-radiation bursts and tunable output field.}---Until now, we have studied the non-Markovian dynamics of the atomic part in the preceding sections. Synergistic effect of emitted photons is another interesting viewpoint to give a more complete dynamical description of the atom-field quantum system. We thus continued by investigating the interference patterns of radiation field, which can be characterized by the bosonic field density distribution (FDD) with a general form $I\propto\bra{\psi(t)}\hat{E}^{\dagger}(x,t)\hat{E}(x,t)\ket{\psi(t)}$, where $\hat{E}(x,t)$ is the electric field operator. When the emitters are initially prepared in the symmetric ($+$) or antisymmetric ($-$) states, the analytical expression of FDD (see Supporting Information S5 for more details) is given by
\begin{align}\label{eq9}
&I_{\pm}(x,t)=\frac{\gamma \pi}{v_{g}^{2}}|\underset{i,j}{\sum}(\pm 1)^{j}(R_{ij}+L_{ij})|^{2},
\end{align}
with
\begin{align}
&R_{ij}= e^{-i\omega_{0}(t-t_{ij})}c_{a}(t-t_{ij})[\Theta(t-t_{ij})-\Theta(-t_{ij})],\nonumber\\
&L_{ij}= e^{-i\omega_{0}(t+t_{ij})}c_{a}(t+t_{ij})[\Theta(t+t_{ij})-\Theta(t_{ij})],\nonumber
\end{align}
where we have defined $t_{ij}\equiv[x+(-1)^{j+1}x_{ai}]/v_{g}$ with $i,\,j=1,\,2$. The influence of atomic geometric structures and retardation $\eta$ on FDD is reflected by $t_{ij}$ and $c_{a}(t)$. One may extract
\begin{figure}
  \centering
  \includegraphics[width=8.0cm]{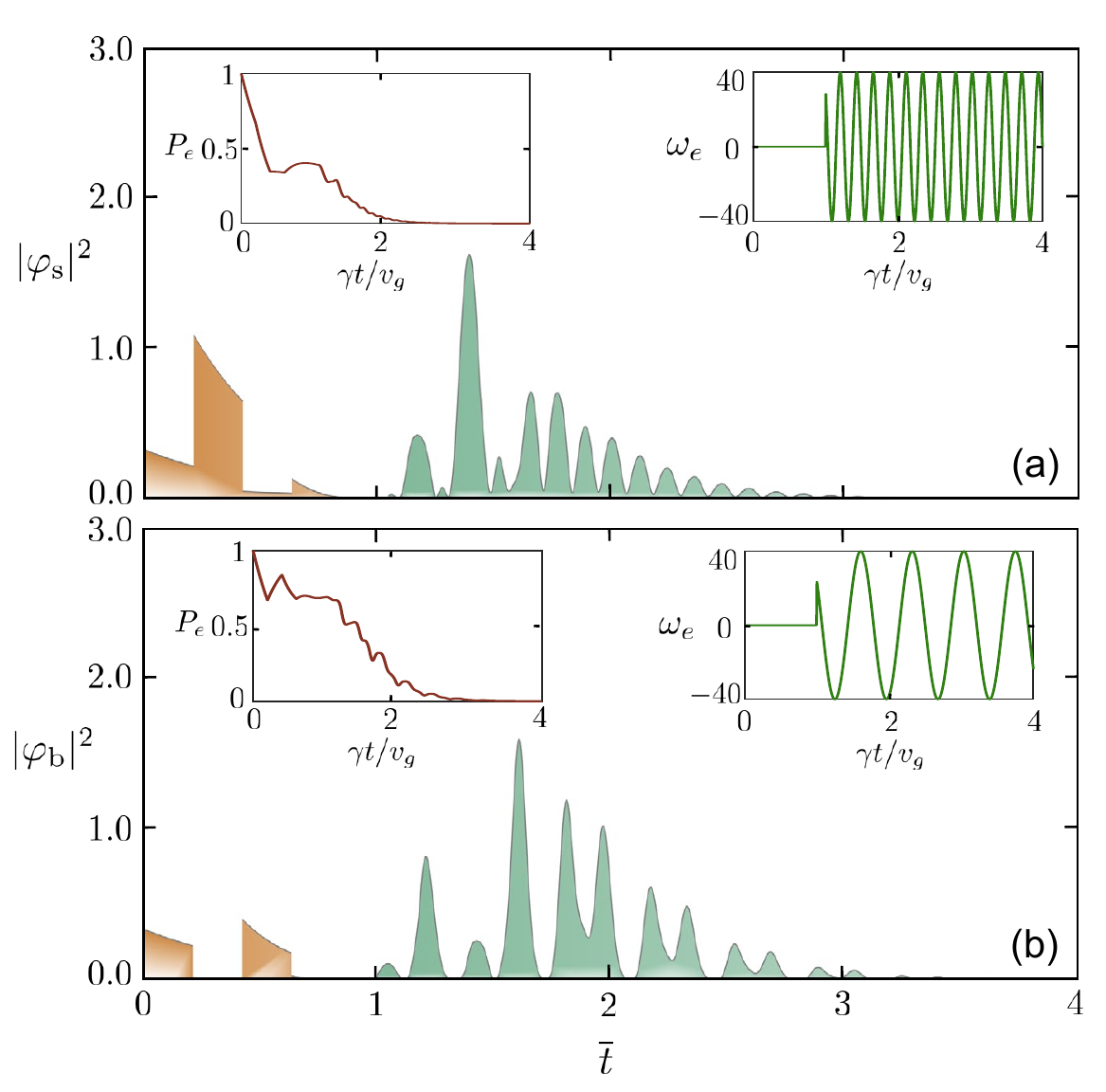}
  \caption{(Color online) Real-time detection of output field intensity of two separate (a) and braided (b) giant atoms. The dynamical evolution process of atomic states and corresponding transition frequency $\omega_{e}$ versus $\gamma t/v_{g}$ are displayed in the insets. The temporal profiles of output field for the first four traveling time $4\Delta t$ (filled by the brown area) in (a) and (b)  are consistent with the cases shown in Figs.\,\ref{fig4}(b) and (d), respectively. After a proper time, we break the parameter condition for dark modes and then the detector receives the optical flow signals again (filled by the dark-green area).}\label{fig5}
\end{figure}
\noindent vivid physical diagrams from Eq.\,(\ref{eq9}) through the unidirectional propagators $L_{ij}$ and $R_{ij}$, which describe the unidirectional radiation processes. More specifically, $R_{11}$ represents the right-moving emission from the leg labeled by $x_{b2}$, and $L_{12}$ represents the left-moving emission from the leg labeled by $x_{a1}$. These radiation fields collide each other to form the interference patterns, which behave superradiant or subradiant.

In Fig.\,\ref{fig4}, we plot the FDD of the bosonic field, where each coupling point leads to a light cone expanding outward as a whole. In the case of superradiance, a collective radiation burst occurs while two cones meet each other, leading to the coherent enhancement. It is shown from Figs.\,\ref{fig4}(a) and (c) that the multi-radiation bursts are eventually formed in a region outside the atomic ensemble in the presence of multiple radiation sources along the 1D waveguide. In another case of subradiance, the local behavior of radiation field can be observed where the trapped photons or phonons bounce back and forth between the connecting points in several specific areas [see Figs.\,\ref{fig4}(b) and (d)].

In the above discussion, we find that certain photons may be trapped in specific regions. Interestingly, the excitations limited in such a photon-emitter bound state can be re-released by applying a time-dependent energy level splitting on the giant atoms. As the giant atoms generally modeled by superconducting qubits, this operation can be realized conveniently in a manner of applying an external bias magnetic field. We determine the dynamical manipulation process of the system by putting a detector at $\overline{x}=x_{b2}+x_{0}\,(x_{0}>0)$. The corresponding output field amplitudes from two separate/braided giant atoms at time $t=\overline{t}+x_{0}/v_{g}$ read
%conveniently
\begin{align}
&|\varphi_{\rm{s}}(\overline{t})|\!=\!\frac{2}{\sqrt{\gamma v_{g}}}[F_{1,3}(\overline{t})\!+\!F_{1,2}(\overline{t})\!+\!F_{2,1}(\overline{t})\!+\!F_{2,0}(\overline{t})],\label{10}\\
&|\varphi_{\rm{b}}(\overline{t})|\!=\!\frac{2}{\sqrt{\gamma v_{g}}}[F_{1,3}(\overline{t})\!+\!F_{1,1}(\overline{t})\!+\!F_{2,2}(\overline{t})\!+\!F_{2,0}(\overline{t})].\label{11}
\end{align}

In Fig.\,\ref{fig5}, we plot the output field intensity $|\varphi_{s/b}(\overline{t})|^{2}$ for photons flowing through $\overline{x}$ over time $\overline{t}$. Atomic ensemble reaches the stable equilibrium by decaying a part of excitations to the waveguide (in a region filled by brown). Just then, we change the atomic energy gap so that $\omega_{0}\Delta t=n\pi$ no longer  holds to meet the conditions of generating system's BIC. Therefore, the detector will detect the optical flow again (in a region filled by dark-green). In a short summary, the number of photons in the cavity formed by the coupling points could be manipulated based on the above procedure.\vspace{0.15cm}

\emph{Summary and outlook.}---In summary, we have studied the non-Markovian collective emission of two giant atoms coupled to a 1D bosonic environment. Based on the analytical and numerical results, we found that this system allows the occurrence of a strongly modified decay process outside the scope of normal super/subradiance, which is originally triggered by both the giant atomic effects and coherent time-delayed feedback. The system also supports the localized states embedded in the continuum, with photons or phonons bouncing back and forth between coupling points in certain selective areas.

This work opens up new eyesight in the non-Markovian dynamical system with multi-delay signals. Possible applications of our results in quantum information science include the realizations of entangled emitter- photon states, dynamically adjustable bound states, long-range interaction between distant emitters, and the time-dependent superradiant laser. It then would be interesting to combine the effects of intricate time-delays and topological or chiral quantum optics\,\cite{P. Lodahl}.

\let\oldaddcontentsline\addcontentsline% Store \addcontentsline
\renewcommand{\addcontentsline}[3]{}% Make \addcontentsline a no-op

\let\addcontentsline\oldaddcontentsline% Restore \addcontentsline

%%%%%%%%%% Merge with supplemental materials %%%%%%%%%%
\onecolumngrid

%%%%%%%%%% Prefix a "S" to all equations, figures, tables and reset the counter %%%%%%%%%%
\newcommand\specialsectioning{\setcounter{secnumdepth}{-2}}
\setcounter{equation}{0} \setcounter{figure}{0}

\setcounter{table}{0}
\renewcommand{\theequation}{S\arabic{equation}}
\renewcommand{\thefigure}{S\arabic{figure}}
\renewcommand{\bibnumfmt}[1]{[S#1]}
\renewcommand{\citenumfont}[1]{S#1}
\renewcommand\thesection{S\arabic{section}}
%%%%%%%%%% Prefix a "S" to all equations, figures, tables and reset the counter %%%%%%%%%%
\renewcommand{\baselinestretch}{1.2}

%\renewcommand{\theequation}{S\arabic{equation}}

%%%%%%%%%%%%%%%%%%%%%%%%%%%%%%%%%%%%%%%%%%%%%%%%%%%%%%%%%%%%%%%%%
\newpage

\setcounter{page}{1}\setcounter{secnumdepth}{3} \makeatletter
\begin{center}
{\Large \textbf{ Supplemental Material for\\
        ``Collective Emission of Giant Atoms in Non-Markovian Regime"}}
\end{center}

\begin{center}
Qing-Yang Qiu$^{1}$, Ying Wu$^{1}$, Xin-You L\"{u}$^{1}$
\end{center}

\begin{minipage}[]{16cm}
\small{\it
	\centering School of Physics, Huazhong University of Science and Technology, Wuhan, 430074, P. R. China \\}
\end{minipage}

\vspace{8mm}

\tableofcontents

%%%%%%%%%%%%%%%%%%%%%%%%%%%%%%%%%%%%%%%%%%%%%%

\section{MULTIDELAY DIFFERENTIAL EQUATION}
We now give a detailed derivation of multi-delay differential equations in the main text. By transforming the total Hamiltonian of the system to the interaction picture and applying the schr$\ddot{o}$dinger equation $i|\dot{\psi}(t)\rangle=\hat{H}_{\rm{int}}\ket{\psi(t)}$, one can obtain the coupled differential equations of the probability amplitudes for the emitter-field system
\begin{align}
&\dot{c}_{m}(t)=-i\underset{n}{\sum}\underset{\alpha=R,L}{\sum}\int_{0}^{\infty}g(\omega)\varphi_{\alpha}(\omega,t)e^{i\epsilon_{\alpha}\omega x_{n}^{m}/v_{g}}e^{-i(\omega-\omega_{0})t}d\omega ,\label{S1}\\
&\dot{\varphi}_{\alpha}(\omega,t)=-i\underset{m}{\sum}\underset{n}{\sum}g(\omega)c_{m}(t)e^{-i\epsilon_{\alpha}\omega x_{n}^{m}/v_{g}}e^{i(\omega-\omega_{0})t},\label{S2}
\end{align}
where $x_{n}^{m}$ represents the position of $n$th connecting point from $m$th giant atom, and $\varphi_{L(R)}(\omega,t)$ represents the amplitude of probability to find a single photon propagating to the left (right) at time $t$ with frequency $\omega$ in the waveguide. Since we consider a pair of two-legged giant atoms coupled with a 1D waveguide, the summation indicators $m,n$ in $x^{m}_{n}$ should be taken 1 or 2 in our case. We proceed by solving $\varphi_{\alpha}(\omega,t)$ formally with the results for the right- and left-going modes
\begin{align}
&\varphi_{R}(\omega,t)=-ig(\omega_{0})\int_{0}^{t}\sum_{m=1,2}\sum_{n=1,2}c_{m}(\tau)e^{-i\omega x_{n}^{m}/v_{g}}e^{i(\omega-\omega_{0})\tau}d\tau,\label{S3}\\
&\varphi_{L}(\omega,t)=-ig(\omega_{0})\int_{0}^{t}\sum_{m=1,2}\sum_{n=1,2}c_{m}(\tau)e^{i\omega x_{n}^{m}/v_{g}}e^{i(\omega-\omega_{0})\tau}d\tau.\label{S4}
\end{align}
The atomic equations of motion (EOMs) can be decoupled from the bosonic modes by substituting Eq.(\ref{S3})-(\ref{S4}) into Eq.(\ref{S1}), i.e.,
\begin{align}\label{S5}
&\dot{c}_{1}(t)=-2\int_{0}^{+\infty}d\omega\int_{0}^{t}d\tau\bigl|g(\omega)\bigr|^{2}(\sum_{m=1,2}e^{i\omega x_{m}^{1}/v_{g}})(\sum_{n=1,2}e^{-i\omega x_{n}^{1}/v_{g}})c_{1}(\tau)e^{-i(\omega-\omega_{0})(t-\tau)}\nonumber\\
&-\int_{0}^{+\infty}d\omega\int_{0}^{t}d\tau\bigl|g(\omega)\bigr|^{2}[(\sum_{m=1,2}e^{i\omega x_{m}^{1}/v_{g}})(\sum_{n=1,2}e^{-i\omega x_{n}^{2}/v_{g}})+(\sum_{m=1,2}e^{-i\omega x_{m}^{1}/v_{g}})(\sum_{n=1,2}e^{i\omega x_{n}^{2}/v_{g}})]c_{2}(\tau)e^{-i(\omega-\omega_{0})(t-\tau)},
\end{align}
\begin{align}\label{S6}
&\dot{c}_{2}(t)=-2\int_{0}^{+\infty}d\omega\int_{0}^{t}d\tau\bigl|g(\omega)\bigr|^{2}(\sum_{m=1,2}e^{i\omega x_{m}^{2}/v_{g}})(\sum_{n=1,2}e^{-i\omega x_{n}^{2}/v_{g}})c_{2}(\tau)e^{-i(\omega-\omega_{0})(t-\tau)}\nonumber\\
&-\int_{0}^{+\infty}d\omega\int_{0}^{t}d\tau\bigl|g(\omega)\bigr|^{2}[(\sum_{m=1,2}e^{i\omega x_{m}^{2}/v_{g}})(\sum_{n=1,2}e^{-i\omega x_{n}^{1}/v_{g}})+(\sum_{m=1,2}e^{-i\omega x_{m}^{2}/v_{g}})(\sum_{n=1,2}e^{i\omega x_{n}^{1}/v_{g}})]c_{1}(\tau)e^{-i(\omega-\omega_{0})(t-\tau)},
\end{align}
where $(\sum_{m=1,2}e^{i\omega x_{m}^{i}/v_{g}})(\sum_{n=1,2}e^{-i\omega x_{n}^{j}/v_{g}})$ depends completely on the specific geometric structure of giant atoms.
\subsection{Long-time limit behavior and the Markovian decay rates for separate atoms}
Here, we first consider two separate giant atoms with two legs interacting with a 1D bosonic modes. In this case, the interference factors in Eq.(\ref{S5})-(\ref{S6}) read
\begin{align}\label{S7}
(\sum_{m=1,2}e^{i\omega x_{m}^{1}/v_{g}})(\sum_{n=1,2}e^{-i\omega x_{n}^{1}/v_{g}})&=2+e^{i\omega\Delta x/v_{g}}+e^{-i\omega\Delta x/v_{g}}\nonumber\\
(\sum_{m=1,2}e^{i\omega x_{m}^{1}/v_{g}})(\sum_{n=1,2}e^{-i\omega x_{n}^{2}/v_{g}})&=2e^{-2i\omega\Delta x/v_{g}}+e^{-i\omega\Delta x/v_{g}}+e^{-3i\omega\Delta x/v_{g}},
\end{align}
where $\Delta x$ denotes the distance between neighboring coupling points. We next assume that $g(\omega)\approx g(\omega_{0})=\sqrt{\gamma/4\pi}$ is constant for all waveguide modes and the resonant frequency  $\omega_{0}$ of two-level approximate emitters is far away from the cut-off frequency of the waveguide. Therefore, combining the Eq.(\ref{S5})-(\ref{S6}) and Eq.(\ref{S7}), we get
\begin{align}\label{S8}
\dot{c}_{m}(t)=&-\gamma c_{m}(t)-\gamma e^{i\varphi}c_{m}(t-\Delta t)\Theta(t-\Delta t)
-\gamma e^{i2\varphi}c_{n}(t-2\Delta t)\Theta(t-2\Delta t)\nonumber\\
&-\frac{\gamma}{2}e^{i\varphi}c_{n}(t-\Delta t)\Theta(t-\Delta t)-\frac{\gamma}{2}e^{i3\varphi}c_{n}(t-3\Delta t)\Theta(t-3\Delta t)
\end{align}
for $m\neq n$. To get a more compact expression, we define $F_{m,n}(t)\equiv\gamma e^{in\varphi}c_{m}(t-n\Delta t)\Theta(t-n\Delta t)$ where $\varphi\equiv\omega_{0}\Delta t=k_{0}\Delta x$, and then Eq.(3) in the main text is obtained. A differential equation with multiple delays like Eq.(\ref{S8}) is mathematically called as a multi-delay differential equation and their solutions are difficult to be expressed by a simple formula.

To give the asymptotic behavior of excitation probability amplitude in the long-time limit and the Markovian collective decay rates, it's useful to transform Eq.(\ref{S8}) into Laplace space, i.e.,
\begin{align}\label{S9}
sc_{m}(s)-\beta_{m}(0)=-\gamma c_{m}(s)-\gamma e^{i\varphi}c_{m}(s)e^{-s\Delta t }-\gamma e^{2i\varphi}c_{n}(s)e^{-2s\Delta t}-\frac{\gamma}{2} e^{i\varphi}c_{n}(s)e^{-s\Delta t}-\frac{\gamma}{2} e^{3i\varphi}c_{n}(s)e^{-3s\Delta t},
\end{align}
where $s$ is the laplace variable, and $\beta_{m}(0)$ is the initial probability amplitude coefficient of $m$th emitter. For illustration purpose, we can rewrite Eq.(\ref{S9}) by considering the atoms are initially prepared in an exchange symmetric state $\ket{+}=\frac{1}{\sqrt{2}}(\left|eg\right\rangle +\left|ge\right\rangle )$ such that
\begin{align}\label{S10}
c_{+}(s)=(s+\gamma+\gamma e^{i\varphi}e^{-s\Delta t}+\gamma e^{i2\varphi}e^{-2s\Delta t}+\frac{\gamma}{2}e^{i\varphi}e^{-s\Delta t}+\frac{\gamma}{2}e^{i3\varphi}e^{-3s\Delta t})^{-1},
\end{align}
where we have defined the probability amplitude $c_{+}=c_{1,+}=c_{2,+}$ for symmetric state. By applying the final value theorem, we can predict the steady-state behaviour in the long-time limit:
\begin{align}\label{S11}
\underset{t\rightarrow\infty}{{\rm{lim}}}c_{+}(t)&=\underset{s\rightarrow0}{{\rm{lim}}}sc_{+}(s)\nonumber\\
&=\underset{s\rightarrow0}{{\rm{lim}}}\frac{s}{s+\gamma+\gamma e^{i\varphi}e^{-s\Delta t}+\gamma e^{i2\varphi}e^{-2s\Delta t}+\frac{\gamma}{2}e^{i\varphi}e^{-s\Delta t}+\frac{\gamma}{2}e^{i3\varphi}e^{-3s\Delta t}}.
\end{align}
It is shown from Eq.\,(\ref{S11}) that $\underset{t\rightarrow\infty}{{\rm{lim}}}c_{+}(t)$ approaches to the asymptotic value $1/(1+\gamma\Delta t)$ when $\varphi=(2n+1)\pi$, which indicates the appearance of subradiant state. On the contrary, the radiant state, i.e., $\underset{t\rightarrow\infty}{{\rm{lim}}}c_{+}(t)\rightarrow 0$, is obtained when $\varphi\neq(2n+1)\pi$.

In the following, we consider the atomic Markovian dynamics by assuming that the coupling points are slightly separated such that $3\gamma\Delta t\ll 1$. And then the analytical expression of effective decay rates in this case could be obtained with the help of Laplace space through which several basic properties can be summarized. In this regime, the Eq.(\ref{S10}) can be further reduced as
\begin{align}\label{S12}
c_{+}(s)=[s(1-2e^{2i\varphi}\eta-\frac{3}{2}e^{i\varphi}\eta-\frac{3}{2}e^{3i\varphi}\eta)+\gamma(1+\frac{3}{2}e^{i\varphi}+e^{2i\varphi}+\frac{1}{2}e^{3i\varphi})]^{-1},
\end{align}
which gives immediately the effective spontaneous emission rate
\begin{align}\label{S13}
\gamma_{+}=\frac{2\gamma(1+\frac{3}{2}e^{i\varphi}+e^{2i\varphi}+\frac{1}{2}e^{3i\varphi})}{1-2e^{2i\varphi}\eta-\frac{3}{2}e^{i\varphi}\eta-\frac{3}{2}e^{3i\varphi}\eta}.
\end{align}
This effective decay rate reveals that the phenomenon of radiation enhancement or inhibition is attributed to the
joint effects of the coherent time-delayed feedback and giant atom interference between the coupling points. In addition, when the separation between legs is very large, i.e., $\eta\gg 1$, then $c_{+}(s)\approx 1/(s+\gamma)$, which means that emitters radiate photons independently, and the decay rate in this case reads $2\gamma$.

Same procedures can be applied when the separate giant atoms are initially prepared in an exchange anti-symmetric state $\ket{-}=\frac{1}{\sqrt{2}}(\left|eg\right\rangle -\left|ge\right\rangle )$. The asymptotic behaviour in the long-time limit can be obtained similarly, from which we find that $\underset{t\rightarrow\infty}{{\rm{lim}}}c_{-}(t)$ approaches to the asymptotic value $1/(1+\gamma\Delta t)$ when $\varphi=(2n+1)\pi$, and approaches to the asymptotic value $1/(1+3\gamma\Delta t)$ when $\varphi=2n\pi$, which indicates the appearance of subradiant state. On the contrary, the radiant state, i.e., $\underset{t\rightarrow\infty}{{\rm{lim}}}c_{-}(t)\rightarrow 0$, is obtained when $\varphi\neq n\pi$. And the corresponding effective spontaneous emission rate is
\begin{align}\label{S14}
\gamma_{-}=\frac{2\gamma(1+\frac{1}{2}e^{i\varphi}-e^{i2\varphi}-\frac{1}{2}e^{i3\varphi})}{1-\frac{1}{2}e^{i\varphi}\eta+2e^{i2\varphi}\eta+\frac{3}{2}e^{i3\varphi}\eta}.
\end{align}
\begin{figure}
  \centering
  \includegraphics[width=17.3cm]{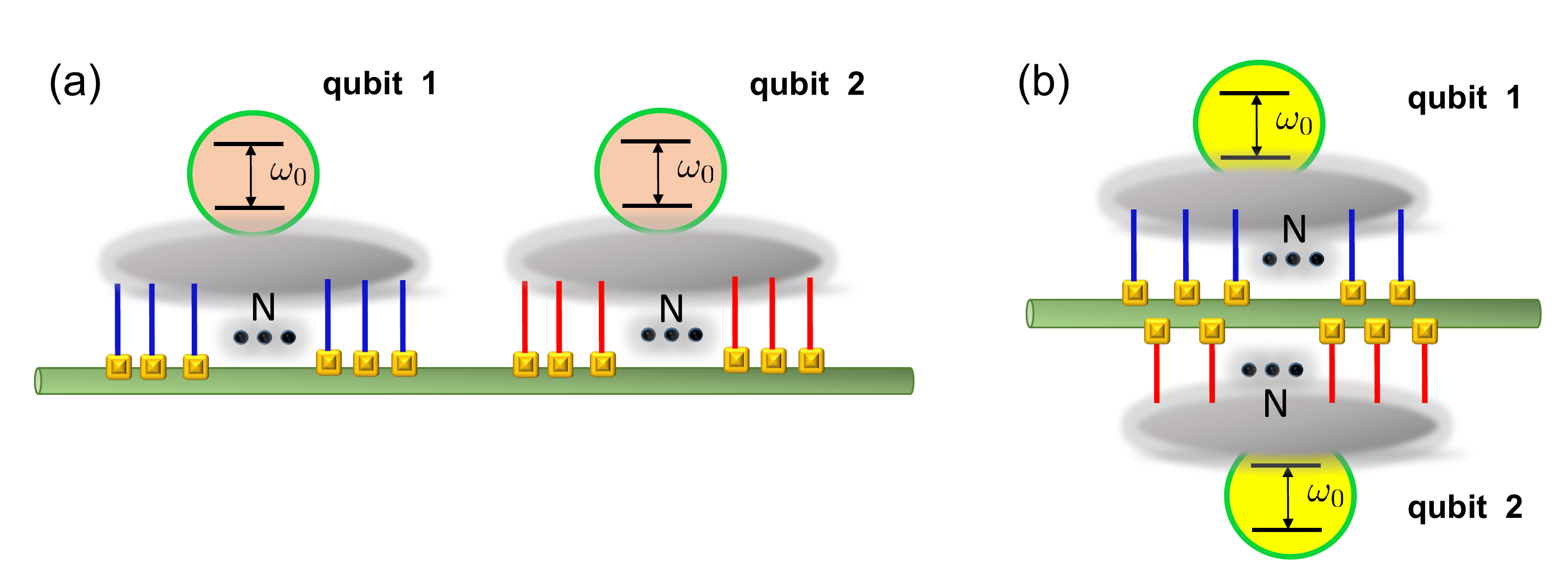}
  \caption{Schematic illustration of two N-legged (a) separate giant atoms and (b) braided giant atoms. }\label{figS1}
\end{figure}

A more general case where two separate N-legged giant atoms coupled to a 1D waveguide as shown in Fig.\,\ref{figS1}(a) has been considered. The corresponding EOMs and effective decay rate of a symmetric state are described by
\begin{align}
\dot{c}_{m}(t)=&-\frac{N\gamma}{2}c_{m}(t)-\sum_{k=1}^{N-1}\gamma(N-k)c_{m}(t-k\Delta t)\Theta(t-k\Delta t)e^{ik\varphi}-\sum_{k=1}^{N}\frac{\gamma k}{2}c_{n}(t-k\Delta t)\Theta(t-k\Delta t)e^{ik\varphi}\nonumber\\
&-\sum_{k=N+1}^{2N-1}\frac{\gamma}{2}(2N-k)c_{n}(t-k\Delta t)\Theta(t-k\Delta t)e^{ik\varphi},\label{S15}\\
\gamma_{+}=&\frac{2\gamma[\frac{N}{2}+\sum_{k=1}^{2N-1}(N-\frac{k}{2})e^{ik\varphi}]}{1-\sum_{k=1}^{2N-1}k\eta(N-\frac{k}{2})e^{ik\varphi}},\label{S16}
\end{align}
for $m\neq n$. Particularly, we obtain a superradiant decay rate $2N^{2}\gamma/[1-\eta(\frac{1}{6}N-\frac{5}{6}N^{2}+N^{3})]$ when we choose the field propagation phase difference $\varphi=2n\pi$ and assume all the coupling points are limited in a narrow area such that $(2N-1)\eta\ll 1$.

\subsection{Long-time limit behavior and the Markovian decay rates for braided atoms}
We now consider two braided giant atoms. The interference factors in Eq.(\ref{S5})-(\ref{S6}) read
\begin{align}\label{S17}
(\sum_{m=1,2}e^{i\omega x_{m}^{1}/v_{g}})(\sum_{n=1,2}e^{-i\omega x_{n}^{1}/v_{g}})&=2+e^{2i\omega\Delta x/v_{g}}+e^{-2i\omega\Delta x/v_{g}},\nonumber\\
(\sum_{m=1,2}e^{i\omega x_{m}^{1}/v_{g}})(\sum_{n=1,2}e^{-i\omega x_{n}^{2}/v_{g}})&=2e^{-i\omega\Delta x/v_{g}}+e^{i\omega\Delta x/v_{g}}+e^{-3i\omega\Delta x/v_{g}}.
\end{align}
Following the similar procedures that have been mentioned above, we obtain the EOMs of the atomic probability amplitude:
\begin{align}\label{S18}
\dot{c}_{m}(t)&=-\gamma c_{m}(t)-\gamma e^{i2\varphi}c_{m}(t-2\Delta t)\Theta(t-2\Delta t)-\frac{3\gamma}{2}e^{i\varphi}c_{n}(t-\Delta t)\Theta(t-\Delta t)-\frac{\gamma}{2}e^{i3\varphi}c_{n}(t-3\Delta t)\Theta(t-3\Delta t)
\end{align}
for $m\neq n$. To give the asymptotic behavior of excitation probability amplitude in the long-time limit and the Markovian collective decay rates, we transform Eq.(\ref{S18}) into Laplace space, i.e.,
\begin{align}\label{S19}
sc_{m}(s)-\beta_{m}(0)=-\gamma c_{m}(s)-\gamma e^{i2\varphi}c_{m}(s)e^{-2s\Delta t}-\frac{3\gamma}{2}e^{i\varphi}c_{n}(s)e^{-s\Delta t}-\frac{\gamma}{2}e^{i3\varphi}c_{n}(s)e^{-3s\Delta t}.
\end{align}

The steady-state behaviour in the long-time limit can be predicted by applying the final value theorem.  We find that a non-radiative state with $\underset{t\rightarrow\infty}{{\rm{lim}}}c_{+}(t)=1/(1+\gamma\Delta t)$ when $\varphi=(2n+1)\pi$ in the case of a symmetric initial state.  When the giant atoms are initially in an anti-symmetric initial state, the amplitude $\underset{t\rightarrow\infty}{{\rm{lim}}}c_{-}(t)=1/(1+\gamma\Delta t)$ for $\varphi=2n\pi$ also indicates that a non-radiative state is obtained. On the contrary, a radiative state state is obtained when $\varphi\neq(2n+1)\pi$ [or $\varphi\neq2n\pi$] in the case of a symmetric [or anti-symmetric] initial state.

To give an analytical understanding in the Markovian regime, we further reduce the Eq.(\ref{S19}) by introducing a very small retardation so that $3\eta\ll 1$. This approximation reveals that the braided atoms share the same dynamical process with the separate giant atoms while atoms are excited symmetrically. When two braided giant atoms are excited anti-symmetrically, the corresponding effective spontaneous emission rate is
\begin{align}\label{S20}
\gamma_{-}=\frac{2\gamma(1-\frac{3}{2}e^{i\varphi}+e^{i2\varphi}-\frac{1}{2}e^{i3\varphi})}{1+\frac{3}{2}e^{i\varphi}\eta-2e^{i2\varphi}\eta+\frac{3}{2}e^{i3\varphi}\eta}.
\end{align}

Here, we also consider a more general case where two braided N-legged giant atoms coupled to a 1D waveguide, as shown in Fig.\,\ref{figS1}(b). The corresponding EOMs and effective decay rate of an anti-symmetric state are described by
\begin{align}
\dot{c}_{m}(t)=&-\frac{N\gamma}{2}c_{m}(t)-\sum_{k=1}^{N-1}\gamma(N-k)c_{m}(t-2k\Delta t)\Theta(t-2k\Delta t)e^{i2k\varphi}\nonumber\\
&-\sum_{k=1}^{N}\frac{\gamma(2N-2k+1)}{2}c_{n}[t-(2k-1)\Delta t]\Theta[t-(2k-1)\Delta t]e^{i(2k-1)\varphi},\label{S21}\\
\gamma_{-}=&\frac{2\gamma[\frac{N}{2}+\sum_{k=1}^{N-1}(N-k)e^{i2k\varphi}-\sum_{k=1}^{N}\frac{(2N-2k+1)}{2}e^{i(2k-1)\varphi}]}{1-\eta\sum_{k=1}^{N-1}2k(N-k)e^{i2k\varphi}+\sum_{k=1}^{N}\frac{(2N-2k+1)}{2}(2k-1)\eta e^{i(2k-1)\varphi}},\label{S22}
\end{align}
for $m\neq n$. Particularly, we obtain a superradiant decay rate $2N^{2}\gamma/[1-\eta\frac{(4N^{3}-N)}{6}]$ when we choose the field propagation phase difference $\varphi=(2n+1)\pi$ and assume all the coupling points are limited in a narrow area such that $(2N-1)\eta\ll 1$.

\section{ANALYTICAL TIME EVOLUTION IN THE NON-MARKOVIAN REGIME}
In the previous discussion, multi-delay differential equations are solved analytically by employing the Markovian approximation, where photons released from a leg can be influenced instantaneously by the other coupling points. Generally speaking, this kind of equation has no fully analytical closed solution and even it's numerical stability problems are still under discussion. To grasp further insights into non-Markov dynamics of giant atoms with arbitrary spatial separation, here, we obtain the analytical description of the system's non-Markovian dynamical evolution via a diagrammatic method\,\cite{SF. Dinc1}.

\begin{figure}
	\centering
	\begin{minipage}{0.9\linewidth}
		\centering
		\includegraphics[width=13.5cm]{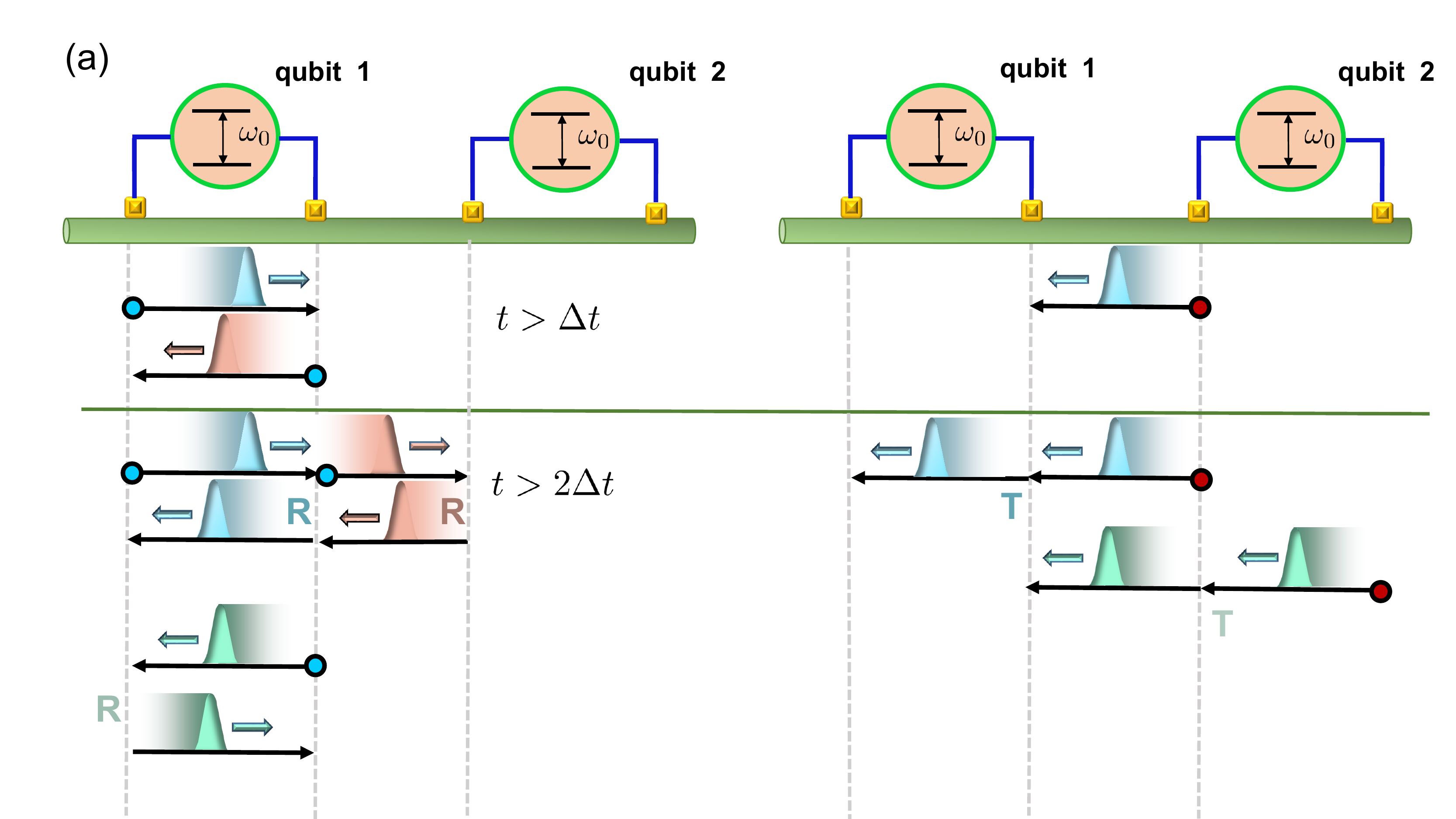}	
	\end{minipage}

	\begin{minipage}{0.9\linewidth}
		\centering
		\includegraphics[width=13.5cm]{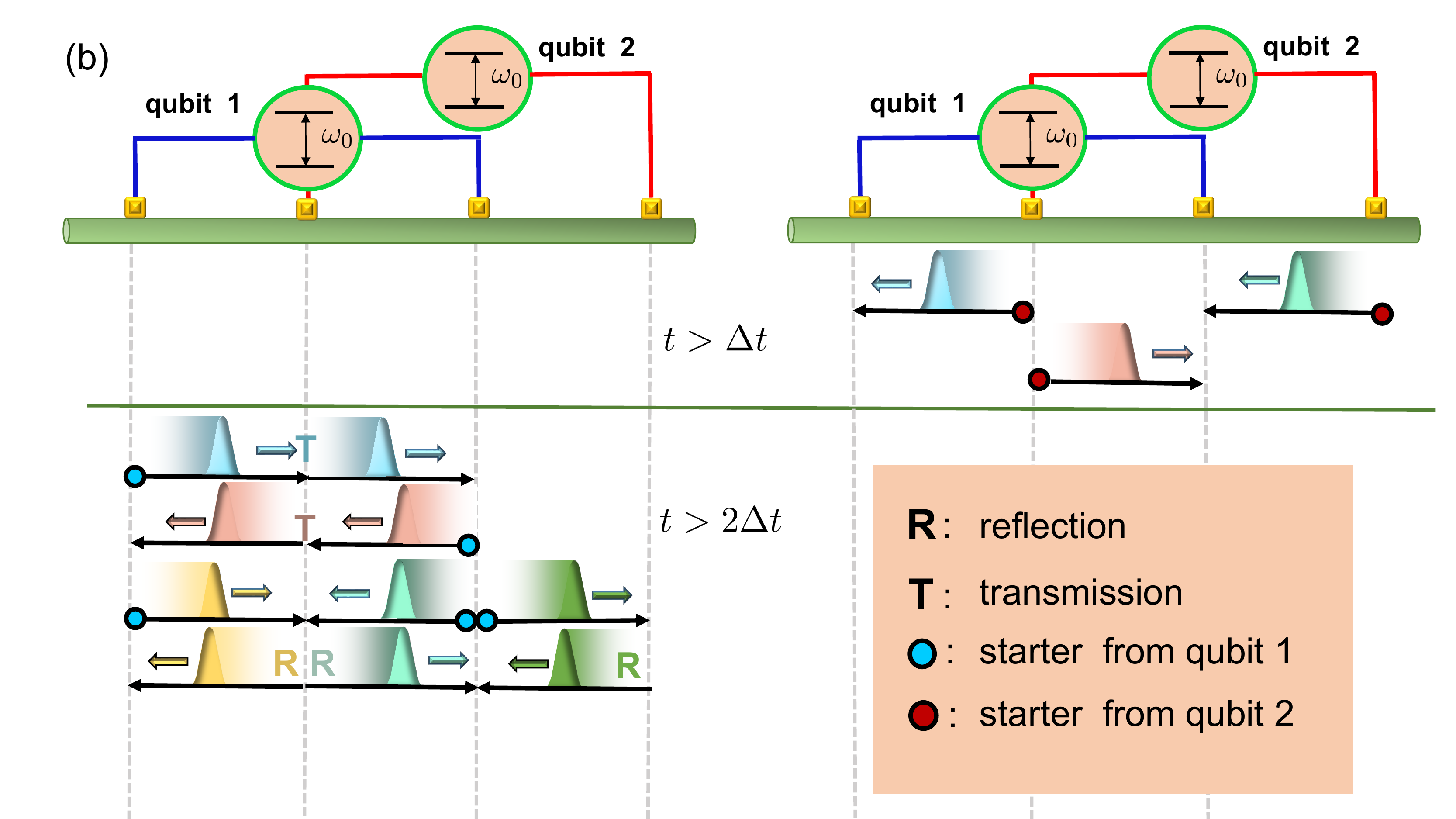}
	\end{minipage}
    \caption{(a) Illustrative diagrams of evolution of the system consisting of two separated giant atoms coupled to a 1D waveguide. They are useful to find the contribution of the excitation amplitude for qubit 1. The right- and left-moving photonic wave packets (the corresponding directions are indicated by arrows) can be reflected (R), transmitted (T) as well as reabsorbed when passing through the coupling points. The upper (or lower) half of figure separated by green line describes the photon propagation routes, that are valid for calculating the evolution of qubit 1 after $t>\Delta t$ (or $t>2\Delta t$). And the left (or right) column depicts the contributory photon transport process where the initial excitation of waveguide is released from qubit 1 (or 2). The wave packets filled by the same color in a certain column belong to a possible dynamical process and each possible photonic trip has only one starter (marked as circle). Diagrams that influence the atomic evolution after $t>3\Delta t$ are no longer displayed here for conciseness. (b) Illustrative diagrams of evolution of the system consisting of two braided giant atoms coupled t a 1D waveguide. }\label{figS2}
\end{figure}

The dynamical process of the system is divided into smaller unit cells that keep track of both the detuned momentum $\Delta_k$ (the detuning energy between light and emitter) and time component $\tau_{l}=t-l\Delta t$ ($l$ is the non-negative integer)\,\cite{SF. Dinc1}. Following the similar spirits in ``collision model", we equivalent the two giant atoms to a set of normal but not dependent emitters\,\cite{SD. Cilluffo}. Under the rule that taking qubit excitation coefficient as finisher, then the excitation amplitude of a considered emitter is described by
\begin{align}\label{S23}
c^{(l)}(t)=\frac{e^{-i\omega_{0}\tau_{l}}}{2\pi}\int_{-\infty}^{\infty}d\Delta_k \,e_{k}\,f_{\rm{in}}^{(l)}e^{-i\Delta_{k}\tau_{l}}c(0),
\end{align}
where $c^{(l)}(t)$ denotes the qubit excitation amplitude and it begins to work after $t>l\Delta t$. The input field $f_{\rm{in}}^{(l)}$ is formed by scattering parameters, e.g., $t_{k}=\frac{\Delta_{k}}{\Delta_{k}+i\gamma}, r_{k}=\frac{-i\gamma}{\Delta_{k}+i\gamma}, e_{k}=\frac{\sqrt{\gamma}}{\Delta_{k}+i\gamma}$. As a result, the atomic probability amplitude can be described as $c(t)=\sum_{l=0}^{\infty}c^{(l)}(t)$.

\subsection{Separate giant atoms}
We now consider the time evolution of two separate giant atoms. Throughout the subsequent calculations, we only focus on the effect of the radiation field on single one qubit, e.g., $c_{a}^{(l)}(t)$. And then the probability amplitudes of collective states can be obtained directly with the help of the correlated properties of the considered symmetric states. The coefficient $c_{a}^{(l)}(t)$ can be solved by finding the input fields $f_{\rm{in}}^{(l)}$. We first consider the separate giant atoms are initially prepared in an exchange symmetric state. In this case, the input fields $f_{\rm{in}}^{(l)}$ can be identified using the excitation's transporting paths as show in Fig.\,\ref{figS2}(a), which give $f_{\rm{in}}^{(0)}=2i/\sqrt{\gamma},f_{\rm{in}}^{(1)}=3e_{k}, f_{\rm{in}}^{(2)}=\frac{5}{2}e_{k}r_{k}+2e_{k}t_{k}$ and $f_{\rm{in}}^{(3)}=\frac{7}{4}e_{k}r_{k}^{2}+e_{k}t_{k}^{2}+4e_{k}t_{k}r_{k}$ for the first four basic traveling time. It is note that the input $f_{\rm{in}}^{(0)}$ is contribute for the spontaneous radiation and works stably throughout the full dynamical process. Analytical atomic evolution can be obtained by substituting these input fields into Eq.(\ref{S23}), and we obtain
\begin{align}\label{S24}
&c_{+}^{(0)}(t)=2\sqrt{2}e^{-(\gamma+i\omega_{0})t}\Theta(t)\tilde{c}_{a}(0)\nonumber\\
&c_{+}^{(1)}(t)=-3\sqrt{2}\gamma e^{-(\gamma+i\omega_{0})(t-\Delta t)}(t-\Delta t)\Theta(t-\Delta t)\tilde{c}_{a}(0)\nonumber\\
&c_{+}^{(2)}(t)=\sqrt{2}e^{-(\gamma+i\omega_{0})(t-2\Delta t)}\frac{1}{2!}\gamma[-4(t-2\Delta t)+\frac{9}{2}\gamma(t-2\Delta t)^{2}]\Theta(t-2\Delta t)\tilde{c}_{a}(0)\nonumber\\
&c_{+}^{(3)}(t)=\sqrt{2}e^{-(\gamma+i\omega_{0})(t-3\Delta t)}[-\gamma(t-3\Delta t)+3\gamma^{2}(t-3\Delta t)^{2}-\frac{9}{8}\gamma^{3}(t-3\Delta t)^{3}]\Theta(t-3\Delta t)\tilde{c}_{a}(0).
\end{align}
Where we have evenly distributed the probability amplitude to each coupling point so that the initial excitation coefficient $\tilde{c}_{a}(0)=c_{a}(0)/2=1/(2\sqrt{2})$. And then the exact atomic probability amplitude can be described as $c_{+}(t)=\sum_{l=0}^{\infty}c_{+}^{(l)}(t)$.

\begin{figure}
  \centering
  \includegraphics[width=17.3cm]{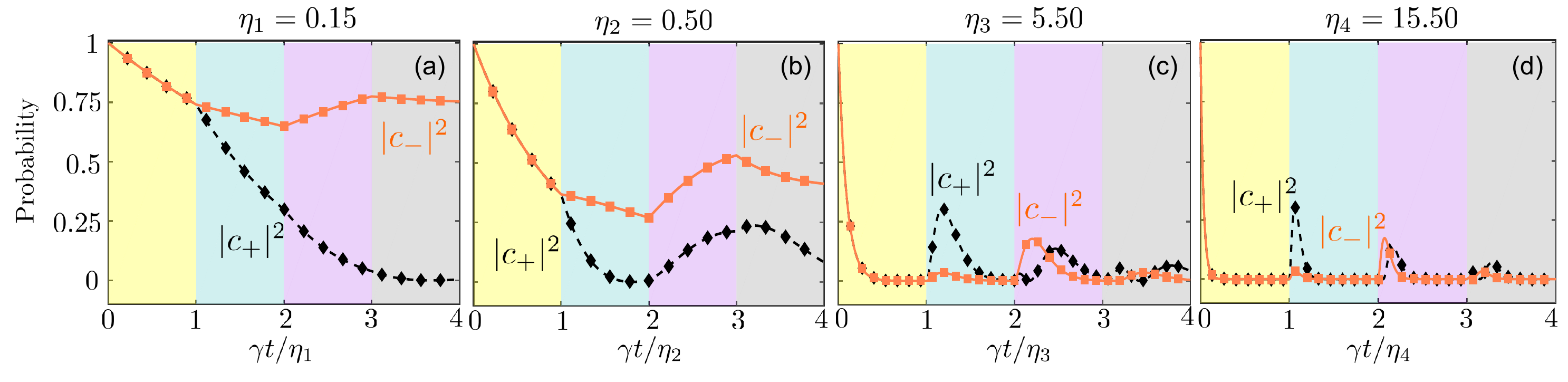}
  \caption{The atomic excitation population $|c(t)|^{2}$ of two separate giant atoms initially prepared in symmetric state (black diamonds and lines) and antisymmetric state (orange squares and lines). The analytical and numerical solutions are shown by the diamonds/squares and lines, respectively. The corresponding parameters are $\varphi=2n\pi$($\varphi=(2n+1)\pi$) for state $|+\rangle$($|-\rangle$). Here the atomic excitation probabilities against time (in units of $\eta/\gamma$) are plotted under increasing values of the retardation $\eta$. (a)$\eta=0.15$. (b)$\eta=0.50$. (c)$\eta=5.50$. (d)$\eta=15.50$.}\label{figS3}
\end{figure}

In the another case when two separated giant atoms initially prepared in an exchange anti-symmetric state, similarly, the input fields for the first four basic traveling time are $f_{\rm{in}}^{(0)}=2i/\sqrt{\gamma},f_{in}^{(1)}=e_{k}, f_{in}^{(2)}=\frac{5}{2}e_{k}r_{k}-2e_{k}t_{k}$ and $f_{in}^{(3)}=\frac{5}{4}e_{k}r_{k}^{2}-e_{k}t_{k}^{2}$, which yields
\begin{align}\label{S25}
&c_{-}^{(0)}(t)=2\sqrt{2}e^{-(\gamma+i\omega_{0})t}\Theta(t)\tilde{c}_{a}(0)\nonumber\\
&c_{-}^{(1)}(t)=-\sqrt{2}\gamma e^{-(\gamma+i\omega_{0})(t-\Delta t)}(t-\Delta t)\Theta(t-\Delta t)\tilde{c}_{a}(0)\nonumber\\
&c_{-}^{(2)}(t)=\sqrt{2}e^{-(\gamma+i\omega_{0})(t-2\Delta t)}\frac{1}{2!}\gamma[4(t-2\Delta t)+\frac{1}{2}\gamma(t-2\Delta t)^{2}]\Theta(t-2\Delta t)\tilde{c}_{a}(0)\nonumber\\
&c_{-}^{(3)}(t)=\sqrt{2}e^{-(\gamma+i\omega_{0})(t-3\Delta t)}[\gamma(t-3\Delta t)-\gamma^{2}(t-3\Delta t)^{2}-\frac{1}{24}\gamma^{3}(t-3\Delta t)^{3}]\Theta(t-3\Delta t)\tilde{c}_{a}(0).
\end{align}
 Then we introduce $D_{ln}(t)\equiv\gamma^{l}(t-n\Delta t)^{l}, K_{l}(t)\equiv\sqrt{2}e^{-(\gamma+i\omega_{0})(t-l\Delta t)}\Theta(t-n\Delta t)c_{a}(0)$ to transform above results into a more compact formalism such that
\begin{align}\label{S26}
c(t)&=\underset{l=0}{\sum^{\infty}}c^{(l)}(t)=\frac{1}{2}\sum^{\infty}_{l=0}K_{l}(t)\sum_{s=0}^{l}D_{ls}(t)f_{s}.
\end{align}
For example, when the two separate giant atoms are prepared initially in a symmetric state, we have
\begin{align}\label{S27}
&c_{+}^{(0)}(t)=\frac{1}{2}K_{0}(t)[2D_{00}(t)]\nonumber\\
&c_{+}^{(1)}(t)=\frac{1}{2}K_{1}(t)[-3D_{11}(t)]\nonumber\\
&c_{+}^{(2)}(t)=\frac{1}{2}K_{2}(t)[-2D_{21}(t)+\frac{9}{4}D_{22}(t)]\nonumber\\
&c_{+}^{(3)}(t)=\frac{1}{2}K_{3}(t)[-D_{31}(t)+3D_{32}(t)-\frac{9}{8}D_{33}(t)].
\end{align}
The comparison results of numerical and analytical calculations are shown in Fig.\,\ref{figS3} under various retardation $\eta$. Here, we only give the atomic evolution behavior for the early four basic time ($0\leq t\leq 4\Delta t$ in our case). For more complete information at longer time, higher-order scattering processes must be taken into account.
\subsection{Braided giant atoms}
We next consider the time evolution of two braided giant atoms.  When the atoms are prepared initially in an exchange symmetric state, the coefficient $c_{a}^{(l)}(t)$ can be solved by finding the input field $f_{\rm{in}}^{(l)}$, which can be identified using the excitation's transporting paths as shown in Fig.\,\ref{figS2}(b). The input fields for the first four basic traveling time are $f_{\rm{in}}^{(0)}=2i/\sqrt{\gamma},f_{\rm{in}}^{(1)}=3e_{k}, f_{\rm{in}}^{(2)}=\frac{5}{2}e_{k}r_{k}+2e_{k}t_{k}$ and $f_{\rm{in}}^{(3)}=\frac{7}{4}e_{k}r_{k}^{2}+e_{k}t_{k}^{2}+4e_{k}t_{k}r_{k}$. Noticed that the given input fields $f_{\rm{in}}$ fully overlap with the case of two separate giant atoms that prepared in a symmetric state. That is to say, both the separate and braided giant atoms follow the the same dynamic process if they are initially prepared in an exchange-symmetric state, which is consistent with the discussion shown in the main text. Analytical atomic evolution can be obtained by substituting these input fields into Eq.(\ref{S23}), and we obtain
\begin{align}\label{S28}
&c_{+}^{(0)}(t)=2\sqrt{2}e^{-(\gamma+i\omega_{0})t}\Theta(t)\tilde{c}_{a}(0)\nonumber\\
&c_{+}^{(1)}(t)=-3\sqrt{2}\gamma e^{-(\gamma+i\omega_{0})(t-\Delta t)}(t-\Delta t)\Theta(t-\Delta t)\tilde{c}_{a}(0)\nonumber\\
&c_{+}^{(2)}(t)=\sqrt{2}e^{-(\gamma+i\omega_{0})(t-2\Delta t)}\frac{1}{2!}\gamma[-4(t-2\Delta t)+\frac{9}{2}\gamma(t-2\Delta t)^{2}]\Theta(t-2\Delta t)\tilde{c}_{a}(0)\nonumber\\
&c_{+}^{(3)}(t)=\sqrt{2}e^{-(\gamma+i\omega_{0})(t-3\Delta t)}[-\gamma(t-3\Delta t)+3\gamma^{2}(t-3\Delta t)^{2}-\frac{9}{8}\gamma^{3}(t-3\Delta t)^{3}]\Theta(t-3\Delta t)\tilde{c}_{a}(0).
\end{align}
If two braided giant atoms are prepared initially in an exchange anti-symmetric state, similarly, the input fields for the first four basic traveling time are $f_{\rm{in}}^{(0)}=2i/\sqrt{\gamma},f_{in}^{(1)}=-3e_{k}, f_{in}^{(2)}=3e_{k}r_{k}+2e_{k}t_{k}$ and $f_{in}^{(3)}=-(\frac{7}{4}e_{k}r_{k}^{2}+e_{k}t_{k}^{2}+4e_{k}t_{k}r_{k})$. Inserting these input fields into Eq.(\ref{S23}), we have
\begin{align}\label{S29}
&c_{-}^{(0)}(t)=2\sqrt{2}e^{-(\gamma+i\omega_{0})t}\Theta(t)\tilde{c}_{a}(0)\nonumber\\
&c_{-}^{(1)}(t)=3\sqrt{2}\gamma e^{-(\gamma+i\omega_{0})(t-\Delta t)}(t-\tau)\Theta(t-\Delta t)\tilde{c}_{a}(0)\nonumber\\
&c_{-}^{(2)}(t)=\sqrt{2}e^{-(\gamma+i\omega_{0})(t-2\Delta t)}\frac{1}{2!}\gamma[-4(t-2\Delta t)+\frac{9}{2}\gamma(t-2\Delta t)^{2}]\Theta(t-2\Delta t)\tilde{c}_{a}(0)\nonumber\\
&c_{-}^{(3)}(t)=\sqrt{2}e^{-(\gamma+i\omega_{0})(t-3\Delta t)}[\gamma(t-3\Delta t)-3\gamma^{2}(t-3\Delta t)^{2}+\frac{9}{8}\gamma^{3}(t-3\Delta t)^{3}]\Theta(t-3\Delta t)\tilde{c}_{a}(0).
\end{align}
Above results can also be transformed into a more compact formalism with the help of the mentioned notation $D_{ls}(t)$ and $K_{l}(t)$.

\section{EXACT COLLECTIVE DECAY RATES IN THE NON-MARKOVIAN REGIME}
\subsection{Separate giant atoms}
We now study the exact collective decay rates of separate giant atoms in the presence of finite delay based on the system Hamiltonian in the real-space ($\hbar =1, v_{g}=1$):
\begin{align}\label{S30}
\hat{H}=&\sum_{m=1,2}\omega_{0}\hat{\sigma}_{+}^{(m)}\hat{\sigma}_{-}^{(m)}+i\int dx[\hat{c}_{L}^{\dagger}(x)\frac{\partial}{\partial x}\hat{c}_{L}(x)-\hat{c}_{R}^{\dagger}(x)\frac{\partial}{\partial x}\hat{c}_{R}(x)]\nonumber\\
&+\int dx\sum_{m=1,2}\sqrt{J_{0}}\delta(x-x_{m})[\sum_{\alpha=R,L}\hat{c}_{\alpha}^{\dagger}(x)\hat{\sigma}^{(1)}_{-}+{\rm{H.c.}}]+\int dx\sum_{m=3,4}\sqrt{J_{0}}\delta(x-x_{m})[\sum_{\alpha=R,L}\hat{c}_{\alpha}^{\dagger}(x)\hat{\sigma}^{(2)}_{-}+{\rm{H.c.}}],
\end{align}
where $\hat{c}^{\dagger}_{L,R}(x)$ is the creation operator of a left- or right- going photon at position $x$, and the position of coupling points $x_{m}$ ordered from left to right, i.e., $x_{1}=x^{1}_{1}, x_{2}=x^{1}_{2}, x_{3}=x^{2}_{1}, x_{4}=x^{2}_{2}$. The light-matter interaction takes the form of delta functions with the coupling strength $\sqrt{J_{0}}$, corresponding to the barriers located at the coupling points. In the single excitation space, we write down the Bathe ansatz for a photon incident from far left:
\begin{align}\label{S31}
|\psi\rangle&=\int dx\,\phi_{R}(x)\hat{c}_{R}^{\dagger}(x)|\emptyset\rangle+\int dx\,\phi_{L}(x)\hat{c}_{L}^{\dagger}(x)|\emptyset\rangle+\xi_{1}\hat{\sigma}^{(1)}_{+}|\emptyset\rangle+\xi_{2}\hat{\sigma}^{(2)}_{+}|\emptyset\rangle,
\end{align}
where $|\emptyset\rangle$ denotes the ground state of the total system in which atoms are all in the lower energy states and waveguide modes are set to be the vacuum state of field. By substituting Eq.(\ref{S30})-(\ref{S31}) into the eigenequation $\hat{H}\ket{\psi}=E\ket{\psi}$, we give the following coupled equation:
\begin{align}
(-i\frac{\partial}{\partial x}-E)\phi_{R}(x)+\sum_{m=1,2}\sqrt{J_{0}}\delta(x-x_{m})\xi_{1}+\sum_{m=3,4}\sqrt{J_{0}}\delta(x-x_{m})\xi_{2}=0 \label{S32}\\
(-i\frac{\partial}{\partial x}-E)\phi_{L}(x)+\sum_{m=1,2}\sqrt{J_{0}}\delta(x-x_{m})\xi_{1}+\sum_{m=3,4}\sqrt{J_{0}}\delta(x-x_{m})\xi_{2}=0 \label{S33}\\
\sum_{m=1,2}\sqrt{J_{0}}[\phi_{R}(x_{m})+\phi_{L}(x_{m})]+(\omega_{0}-E)\xi_{1}=0 \label{S34} \\
\sum_{m=3,4}\sqrt{J_{0}}[\phi_{R}(x_{m})+\phi_{L}(x_{m})]+(\omega_{0}-E)\xi_{2}=0, \label{S35}
\end{align}
where $\phi_{R}(x)$ and $\phi_{L}(x)$ take the form of
\begin{align}
\phi_{R}(x)&=e^{ikx}[\Theta(x_{1}-x)+\sum_{m=1}^{3}t_{m}\Theta(x-x_{m})\Theta(x_{m+1}-x)+t\Theta(x-x_{4})] \label{S36} \\
\phi_{L}(x)&=e^{-ikx}[r\Theta(x_{1}-x)+\sum_{m=1}^{3}r_{m}\Theta(x-x_{m})\Theta(x_{m+1}-x)]. \label{S37}
\end{align}
Substituting the Eqs.(\ref{S36})-(\ref{S37}) into Eqs.(\ref{S32})-(\ref{S35}), we can find the external transmission and reflection coefficients as follows:
\begin{align}
t&=-\frac{e^{-2ik\Delta x}[(-1+e^{2ik\Delta x})J_{0}-ie^{-ik\Delta x}\Delta_k]^{2}}{(1+e^{ik\Delta x})^{2}(-4+e^{2ik\Delta x}+2e^{3ik\Delta x}+e^{4ik\Delta x})J_{0}^{2}+4i(1+e^{ik\Delta x})J_{0}\Delta_k+\Delta_k^{2}} \label{S38} \\
r&=-\frac{(1+e^{ik\Delta x})^{2}J_{0}[2J_{0}(-1-e^{ik\Delta x}+e^{3ik\Delta x}+e^{4ik\Delta x})+i\Delta_k(1+e^{4ik\Delta x})]}{(1+e^{ik\Delta x})^{2}(-4+e^{2ik\Delta x}+2e^{3ik\Delta x}+e^{4ik\Delta x})J_{0}^{2}+4i(1+e^{ik\Delta x})J_{0}\Delta_k+\Delta_k^{2}}, \label{S39}
\end{align}
where $\Delta_k\equiv(k-\omega_{0})$ is the detuning between the photon energy and the atomic energy separation. By setting the denominator of scattering parameters to be zero, we obtain the relevant characteristic equation:
\begin{align}\label{S40}
&J_{0}^{2}[e^{6i(\Delta_k+\omega_{0})\Delta x}+4e^{5i(\Delta_k+\omega_{0})\Delta x}+6e^{4i(\Delta_k+\omega_{0})\Delta x}+4e^{3i(\Delta_k+\omega_{0})\Delta x}-3e^{2i(\Delta_k+\omega_{0})\Delta x}] \nonumber\\
&+(4iJ_{0}\Delta_k-8J_{0}^{2})e^{6i(\Delta_k+\omega_{0})\Delta x}+4iJ_{0}\Delta_k+(\Delta_k^{2}-4J_{0}^{2})=0.
\end{align}
Under the Markovian approximation, we can reduce Eq.(\ref{S40}) to a polynomial characteristic equation ($\tilde{\Delta}_{k}\equiv \Delta_{k}/J_{0}$):
\begin{align}\label{S41}
e^{6i\omega_{0}\Delta x}+4e^{5i\omega_{0}\Delta x}+6e^{4i\omega_{0}\Delta x}+4e^{3i\omega_{0}\Delta x}-3e^{2i\omega_{0}\Delta x}+(4i\tilde{\Delta}_k-8)e^{i\omega_{0}\Delta x}+4i\tilde{\Delta}_k+(\tilde{\Delta}_k^{2}-4)=0.
\end{align}
Poles of this polynomial equation are given by
\begin{align}
\Delta_{k_{-}}&=i(-2-e^{i\varphi}+2e^{2i\varphi}+e^{3i\varphi})J_{0} \label{S42},\\
\Delta_{k_{+}}&=-i(2+3e^{i\varphi}+2e^{2i\varphi}+e^{3i\varphi})J_{0},\label{S43}
\end{align}
then the relevant roots give the collective
decay rates via the Wick-like rotation $\Gamma = 2i\Delta_k$. Thus the analytical collective decay rates of separate giant atoms read
\begin{align}
\Gamma_{-,M}&=(2+e^{i\varphi}-2e^{2i\varphi}-e^{3i\varphi})\gamma \label{S44} \\
\Gamma_{+,M}&=(2+3e^{i\varphi}+2e^{2i\varphi}+e^{3i\varphi})\gamma,\label{S45}
\end{align}
where $\gamma=2J_{0}$ and $\Gamma_{+/-,M}$ is the symmetric/anti-symmetric collective decay rate in the Markov regime. If we set $\varphi=2n\pi$, we find the symmetric (anti-symmetric) mode turns into superradiant (subradiant). However, both the two collective decay rates become zero if we set $\varphi=(2n+1)\pi $. The given analytical effective decay rates in the Markovian limit is reminiscent of the recent findings using the SLH formalism\,\cite{SA. F. Kockum1}.

For the non-Markovian regime, nonlinear effect of transcendental equation in Eq.(\ref{S40}) becomes more and more obvious such that numerical approaches must be utilized. We calculate the collective decay rates by solving Eq.(\ref{S40}) iteratively based on the similar strategy in \cite{SF. Dinc2}. After which the space of collective decay rates is divided into two subspaces called symmetric and anti-symmetric collective decay rates. Noticed that the poles $\Delta_{k_{\pm}}$ can be distinguished due to the fact that the symmetric (anti-symmetric) collective decay rate only contribute to the evolution of symmetric (anti-symmetric) collective state.

The single photon wavepacket, confined between the outmost legs of atomic ensemble, has endless possible round trips before escaping completely from this atomic area. Each potential release path of the photon wavepacket under the delayed feedback mechanism will contribute a physical solution for Eq.(\ref{S40}). In this situation, the ``two-pole" approximation\,\cite{SH. Zheng} is a good choice to pick only symmetric and antisymmetric collective decay rates, on account of extremely weak population of all other collective modes.
\subsection{Braided giant atoms}
The real-space Hamiltonian of two braided giant atoms is described by ($\hbar =1 , v_{g}=1$)
\begin{align}\label{S46}
\hat{H}=&\sum_{n=1,2}\omega_{0}\hat{\sigma}_{+}^{(m)}\hat{\sigma}_{-}^{(m)}+i\int dx[\hat{c}_{L}^{\dagger}(x)\frac{\partial}{\partial x}\hat{c}_{L}(x)-\hat{c}_{R}^{\dagger}(x)\frac{\partial}{\partial x}\hat{c}_{R}(x)]\nonumber\\
&+\int dx\sum_{m=1,3}\sqrt{J_{0}}\delta(x-x_{m})[\sum_{\alpha=R,L}\hat{c}_{\alpha}^{\dagger}(x)\hat{\sigma}^{(1)}_{-}+{\rm{H.c.}}]+\int dx\sum_{m=2,4}\sqrt{J_{0}}\delta(x-x_{m})[\sum_{\alpha=R,L}\hat{c}_{\alpha}^{\dagger}(x)\hat{\sigma}^{(2)}_{-}+{\rm{H.c.}}].
\end{align}
The position of coupling points $x_{m}$ also ordered from left to right, i.e., $x_{1}=x^{1}_{1}, x_{2}=x^{2}_{1}, x_{3}=x^{1}_{2}, x_{4}=x^{2}_{2}$. Following the similar treatments, we obtained the expressions of the single-photon transmission and reflection amplitudes:
\begin{align}
t&=\frac{e^{-2ik\Delta x}[(-1+e^{2ik\Delta x})^{2}J_{0}^{2}+2i(-1+e^{4ik\Delta x})J_{0}\Delta_k+e^{2ik\Delta x}\Delta_k^{2}]}{(-4+e^{2ik\Delta x}+2e^{4ik\Delta x}+e^{6ik\Delta x})J_{0}^{2}+4i(1+e^{2ik\Delta x})J_{0}\Delta_k+\Delta_k^{2}}, \label{S47}\\
r&=-\frac{(1+e^{2ik\Delta x})^{2}J_{0}[2J_{0}(-1+e^{2ik\Delta x})+i\Delta_k(1+e^{2ik\Delta x})]}{(-4+e^{2ik\Delta x}+2e^{4ik\Delta x}+e^{6ik\Delta x})J_{0}^{2}+4i(1+e^{2ik\Delta x})J_{0}\Delta_k+\Delta_k^{2}}.\label{S48}
\end{align}
From which we can arrive at the characteristic equation
\begin{align}\label{S49}
e^{6i(\Delta_k+\omega_{0})L}J_{0}^{2}+2e^{4i(\Delta_k+\omega_{0})L}J_{0}^{2}+J_{0}(J_{0}+4i\Delta_k)e^{2i(\Delta _k+\omega_{0})L}-4J_{0}^{2}+4iJ_{0}\Delta_k+\Delta_k^{2}=0.
\end{align}
Similarly, several basic properties under the Markovian approximation can be easily exposed by setting $\Delta_k$ in exponent part to be zero, after which we find
\begin{align}
\Gamma_{+,M}&=(2+2e^{2i\varphi}+3e^{i\varphi}+e^{3i\varphi})\gamma, \label{S50}\\
\Gamma_{-,M}&=(2+2e^{2i\varphi}-3e^{i\varphi}-e^{3i\varphi})\gamma.\label{S51}
\end{align}
If we set $\varphi=2n\pi$, we find the symmetric mode becomes superradiant ($\Gamma_{+}=8\gamma)$, and anti-symmetric mode becomes subradiant ($\Gamma_{-}=0$). This correspondence will be interchanged if we set $\varphi=(2n+1)\pi$. The results in the Markovian framework are consistent with the previous discussion [see Eq.(\ref{S20})]. Analogously, the exact non-Markovian decay rates are solved numerically as shown in Fig.2(c) in the main text.

\section{BOUND STATES IN THE CONTINUUM}
\subsection{Braided giant atoms}
We first calculate the BICs of the system where two braided giant atoms are trapped along a 1D waveguide. The considered model Hamiltonian in the schr$\ddot{o}$dinger picture reads
\begin{align}\label{S52}
\hat{H}=\sum_{m=1,2}\omega_{0}\hat{\sigma}_{+}^{(m)}\hat{\sigma}_{-}^{(m)}+\int_{-\infty}^{+\infty}dk \,\omega_{k}\hat{b}^{\dagger}(k)\hat{b}(k)+\sum_{m=1,2}\int_{-\infty}^{+\infty}dk[g(k)\hat{\sigma}_{+}^{(m)}\hat{b}(k)(e^{ikx_{1}^{m}}+e^{ikx_{2}^{m}})+{\rm{H.c.}}],
\end{align}
where $\hat{b}(k)$ is the annihilation operator of the waveguide photons with momentum $k$. A general wave function of localized states is of the form
\begin{align}\label{S53}
\ket{\psi}_{\rm{BIC}}=\varepsilon_{1}|e_{1}g_{2}\rangle\otimes|{\rm{vac}}\rangle+\varepsilon_{2}|g_{1}e_{2}\rangle\otimes\ket{{\rm{vac}}}+\int_{-\infty}^{+\infty}dk\varphi_{k}\hat{b}^{\dagger}(k)\ket{g_{1},g_{2},\{0\}},
\end{align}
where $\varepsilon_{1,2}$ is the excitation probability amplitude for the first or second giant atom, and $\varphi_{k}$ is the field amplitude in the momentum space. Inserting Eqs.(\ref{S52}) and (\ref{S53}) into time-independent schr$\ddot{o}$dinger equation, we obtain
\begin{align}
&\dot{\varepsilon}_{1}(t)=-i\omega_{0}\varepsilon_{1}(t)-i\int_{-\infty}^{+\infty}dk\,g(k)\varphi_{k}(t)(e^{ikx_{1}^{1}}+e^{ikx_{2}^{1}})\label{S54},\\
&\dot{\varepsilon}_{2}(t)=-i\omega_{0}\varepsilon_{2}(t)-i\int_{-\infty}^{+\infty}dk\,g(k)\varphi_{k}(t)(e^{ikx_{1}^{2}}+e^{ikx_{2}^{2}})\label{S55},\\
&\dot{\varphi}_{k}(t)=-i\omega_{k}\varphi_{k}(t)-ig^{*}(k)[\varepsilon_{1}(t)(e^{-ikx_{1}^{1}}+e^{-ikx_{2}^{1}})+\varepsilon_{2}(t)(e^{-ikx_{1}^{2}}+e^{-ikx_{2}^{2}})].\label{S56}
\end{align}
We proceed by replacing $\partial t\longrightarrow-iE$ and defining $q\equiv(E-E_{0})/v_{g}$, where $E_{0}$ is the resonant energy of atoms\,\cite{ST. Tufarelli}. Therefore the above coupling equations of emitter-field steady amplitudes can be transformed into
\begin{align}
&q\varepsilon_{1}=\frac{1}{v_{g}}\int_{-\infty}^{+\infty}dk\,g(k)\varphi_{k}(e^{ikx_{1}^{1}}+e^{ikx_{2}^{1}})\label{S57},\\
&q\varepsilon_{2}=\frac{1}{v_{g}}\int_{-\infty}^{+\infty}dk\,g(k)\varphi_{k}(e^{ikx_{1}^{2}}+e^{ikx_{2}^{2}})\label{S58},\\
&[q-(k-k_{0})]\varphi_{k}=\frac{g^{*}(k)}{v_{g}}[\varepsilon_{1}(t)(e^{-ikx_{1}^{1}}+e^{-ikx_{2}^{1}})+\varepsilon_{2}(t)(e^{-ikx_{1}^{2}}+e^{-ikx_{2}^{2}})].\label{S59}
\end{align}
The normalization condition must be satisfied if the BICs are existent, i.e., $|\varepsilon_{1}|^{2}+|\varepsilon_{2}|^{2}+\int_{-\infty}^{+\infty}dk|\varphi_{k}|^{2}=1$, and more concretely,
\begin{align}\label{S60}
&|\varepsilon_{1}|^{2}+|\varepsilon_{2}|^{2}+\int_{-\infty}^{+\infty}dk\frac{|g(k)|^{2}}{v_{g}^{2}[q-(k-k_{0})]^{2}}[\,|\varepsilon_{1}|^{2}(2+e^{ik(x_{1}^{1}-x_{2}^{1})}+e^{ik(x_{2}^{1}-x_{1}^{1})})+|\varepsilon_{2}|^{2}(2+e^{ik(x_{1}^{2}-x_{2}^{2})}+e^{ik(x_{2}^{2}-x_{1}^{2})})\nonumber\\
&+\varepsilon_{1}^{*}\varepsilon_{2}(e^{ikx_{1}^{1}}+e^{ikx_{2}^{1}})(e^{-ikx_{1}^{2}}+e^{-ikx_{2}^{2}})+\varepsilon_{2}^{*}\varepsilon_{1}(e^{ikx_{1}^{1}}+e^{ikx_{2}^{1}})(e^{-ikx_{1}^{2}}+e^{-ikx_{2}^{2}})\,]=1.
\end{align}
For braided giant atoms, spatial configuration of coupling points fulfills $x_{2}^{2}-x_{2}^{1}=x_{1}^{2}-x_{1}^{1}=x_{2}^{1}-x_{1}^{2}=\Delta x\equiv d$. Assuming that the atomic initial state is set to be exchange anti-symmetric state (means that $\varepsilon=\varepsilon_{1}=-\varepsilon_{2}$ ), then Eq.(\ref{S60}) can be simplified as
\begin{align}\label{S61}
|\varepsilon|^{2}\{2+\frac{|g(k_{0})|^{2}}{v_{g}^{2}}\int_{-\infty}^{+\infty}dk\frac{8\cos^{2}kd-6 \cos kd-2{ \cos 3kd}}{[k-(q+k_{0})]^{2}}\}=1.
\end{align}
 Here, we have assumed a flat spectral density of the waveguide modes around the resonance of the qubits such that $|g(k)|^{2}=|g(k_{0})|^{2}=(\gamma v_{g})/(2\pi)$ . Under this approximation, one may find $|\varepsilon_{b}|^{2}=|\varepsilon|^{2}=\frac{1}{2(1+\gamma\Delta t)}$ by calculating the above integral over k through the standard contour-integration methods. And the corresponding field amplitude reads $\varphi_{k}=\varepsilon_{b}\sqrt{\frac{\gamma v_{g}}{2\pi}}\frac{-1}{\omega-\omega_{0}}2i(\sin\frac{3}{2}kd-\sin\frac{1}{2}kd)$. Hence, the $\rm{BIC}$ of this system is described by
\begin{align}\label{S62}
\!\!\ket{\rm{BIC}}=\frac{1}{\sqrt{1+\gamma\varDelta t}}\{\frac{1}{\sqrt{2}}(|eg\rangle-|ge\rangle)\otimes|{\rm{vac}}\rangle-i\sqrt{\frac{\gamma v_{g}}{2\pi}}\int_{0}^{+\infty}dk\frac{1}{\omega-\omega_{0}}(\sin\frac{3}{2}kd- \sin\frac{1}{2}kd)[\hat{a}_{R}^{\dagger}(\omega)-\hat{a}_{L}^{\dagger}(\omega)]|\emptyset\rangle.
\end{align}
For illustration purposes, let us set $q=0$,i.e., $E=\omega_{0}$. We can find a value $k=k_{0}+q$ that makes the right hand side of Eq.(\ref{S59}) vanish, i.e., $ \sin\frac{3}{2}k_{0}d-\sin\frac{1}{2}k_{0}d=0$. In other words, a BIC resonant with emitter's transition frequency is generated. As a result, the generation of BIC for two braided giant atoms occurs in the case of $\varphi=2n\pi$, but not $(2n+1)\pi$, which is consistent with the previous discussions.

\subsection{Separate giant atoms}
We next calculate the bound states in the continuum of the system where two separate giant atoms are trapped along a 1D waveguide. Spatial configuration of legs in this case satisfies the relation $x_{2}^{2}-x_{1}^{2}=x_{1}^{2}-x_{2}^{1}=x_{2}^{1}-x_{1}^{1}\equiv d$. If two giant atoms are prepared initially in an anti-symmetric state $|-\rangle$, then the similar normalization condition is given by
\begin{align}\label{S63}
|\varepsilon|^{2}\{2+\frac{|g(k_{0})|^{2}}{v_{g}^{2}}\int_{-\infty}^{+\infty}dk\frac{8{\sin^{2}kd}+4\sin 2kd\, \sin kd}{[k-(q+k_{0})]^{2}}\}&=1.
\end{align}
Calculating the above integral over $k$ through the standard contour-integration methods, we find that atomic excitation probability amplitude is $|\varepsilon_{b}|^{2}=\frac{1}{2(1+3\gamma\varDelta t)}$ and the relevant field amplitude in the $k$ space has the form of $\varphi_{k}=\varepsilon_{b}\sqrt{\frac{\gamma v_{g}}{2\pi}}\frac{-2i}{\omega-\omega_{0}}( \sin\frac{3}{2}kd+\sin\frac{1}{2}dk)$, which yields immediately the system's BIC:
\begin{align}\label{S64}
\ket{\rm{BIC}}=\frac{1}{\sqrt{1+3\gamma\varDelta t}}\{\frac{1}{\sqrt{2}}(|eg\rangle-|ge\rangle)\otimes|{\rm{vac}}\rangle-i\sqrt{\frac{\gamma v_{g}}{2\pi}}\int_{0}^{+\infty}dk\frac{1}{\omega-\omega_{0}}(\sin\frac{3}{2}kd+ \sin\frac{1}{2}kd)(\hat{a}_{R}^{\dagger}(\omega)-\hat{a}_{L}^{\dagger}(\omega))|\emptyset\rangle.
\end{align}
As we can see, the relation ${\rm sin}\frac{3}{2}k_{0}d+{\rm sin}\frac{1}{2}k_{0}d=0$ must be satisfied to generate a BIC that resonant with emitter's transition frequency. Consequently, the generation of BICs for two separate giant atoms occurs in both the case of $\varphi=2n\pi$ and $\varphi=(2n+1)\pi$.

\section{FIELD INTENSITY DISTRIBUTION}
\subsection{Field intensity distribution for a symmetric state}
The intensity of the field emitted from the atomic ensemble as a function of position and time can be expressed as\,\cite{SK. Sinha}
\begin{align}\label{S65}
I(x,t)=I_{0}\bra{\psi(t)}\hat{E}^{\dagger}(x,t)\hat{E}(x,t)\ket{\psi(t)},
\end{align}
where $\hat{E}(x,t)=\frac{1}{v_{g}}\int_{0}^{+\infty}d\omega\,\varepsilon_{\omega}[e^{i\omega x/v_{g}}\hat{a}_{R}(\omega)+e^{-i\omega x/v_{g}}\hat{a}_{L}(\omega)]e^{i\omega t}$ is the electric field operator at position $x$ and time $t$. Here, we assume $\varepsilon_{\omega}=\varepsilon_{\omega_{0}}$ to be constant for all waveguide modes and the initial state of giant atoms is an exchange symmetric state. Using the fact that $\varphi_{R}(\omega,t)=\varphi_{L}(\omega,t),c_{1}(t)=c_{2}(t)$, one can obtain the field intensity distribution as follows:
\begin{align}\label{S66}
I_{+}(x,t)&=\frac{1}{v_{g}^{2}} \!\bra{\psi(t)}\!\iint_{0}^{+\infty}\!d\omega_{1}d\omega_{2}[e^{-i\omega_{1}x/v_{g}}\hat{a}_{R}^{\dagger}(\omega_{1})\!+\!e^{i\omega_{1}x/v_{g}}\hat{a}_{L}^{\dagger}(\omega_{1})]e^{-i\omega_{1}t}[e^{i\omega_{2}x/v_{g}}\hat{a}_{R}(\omega_{2})\!+\!e^{-i\omega_{2}x/v_{g}}\hat{a}_{L}(\omega_{2})]e^{i\omega_{2}t}\!\ket{\psi(t)}\! \nonumber\\
&=\frac{4}{v_{g}^{2}}\int_{0}^{+\infty}d\omega_{1}\int_{0}^{+\infty}d\omega_{2}\,\varphi_{R}^{*}(\omega_{1},t)\varphi_{R}(\omega_{2},t)\cos(\omega_{1}x/v_{g})\cos(\omega_{2}x/v_{g})e^{-i(\omega_{1}-\omega_{2})t}\nonumber\\ &=\frac{4}{v_{g}^{2}}\left|\int_{0}^{+\infty}d\omega\, \varphi_{R}(\omega,t)\cos(\omega x/v_{g})e^{-i\omega t}\right|^{2}.
\end{align}
Noticed that $I_{+}(x,t)$ has been normalized by the normalization constant $I_{0}\varepsilon_{\omega_{0}}^{2}$ such that $I_{+}(x,t)\rightarrow I_{+}(x,t)/(I_{0}\varepsilon_{\omega_{0}}^{2})$. Substituting the field excitation amplitude $\varphi_{R}(\omega,t)$ in Eq.(\ref{S3}) into Eq.(\ref{S66}), we obtain
\begin{align}\label{S67}
I_{+}(x,t)&=\frac{4\left|g(\omega_{0})\right|^{2}}{v_{g}^{2}}\left|\int_{0}^{+\infty}d\omega\int_{0}^{t}d\tau\sum_{m,n=1,2}c_{m}(\tau)e^{-i\omega x_{n}^{m}/v_{g}}e^{i(\omega-\omega_{0})\tau}\cos(\omega x/v_{g})e^{-i\omega t}\right|^{2}\nonumber\\
&=\frac{\gamma}{\pi v_{g}^{2}}\left|\int_{0}^{+\infty}d\omega\int_{0}^{t}d\tau \cos(\omega x/v_{g})e^{-i\omega t}(e^{-i\omega x_{1}^{1}}+e^{-i\omega x_{2}^{1}}+e^{-i\omega x_{1}^{2}}+e^{-i\omega x_{2}^{2}})c_{1}(\tau)e^{i(\omega-\omega_{0})\tau}\right|^{2}.
\end{align}
Taking the center of the two giant atoms as the origin of coordinates such that $x_{1}^{1}=-x_{2}^{2},x_{2}^{1}=-x_{1}^{2}$ (noticed that this relational expression is appropriate for both the separate and braided giant atoms), then we have
\begin{align}\label{S68}
I_{+}(x,t)=&\frac{4\gamma}{\pi v_{g}^{2}}\left|\int_{0}^{t}d\tau c_{1}(\tau)e^{-i\omega_{0}\tau}\int_{0}^{+\infty}d\omega e^{-i\omega(t-\tau)}[\cos(\omega x/v_{g})\cos(\omega x_{1}^{1}/v_{g})+\cos(\omega x/v_{g})\cos(\omega x_{2}^{1}/v_{g})]\right|^{2}\nonumber\\
=&\frac{\gamma\pi}{v_{g}^{2}} \bigg|\int_{0}^{t}d\tau c_{1}(\tau)e^{-i\omega_{0}\tau}\{\delta[\tau-(t-(x+x_{1}^{1})/v_{g})]+\delta[\tau-(t-(x-x_{1}^{1})/v_{g})]+\delta[\tau-(t+(x-x_{1}^{1})/v_{g})]\nonumber\\
&+\delta[\tau-(t+(x+x_{1}^{1})/v_{g})]+\delta[\tau-(t-(x+x_{2}^{1})/v_{g})]+\delta[\tau-(t-(x-x_{2}^{1})/v_{g})]
+\delta[\tau-(t+(x-x_{2}^{1})/v_{g})]\nonumber\\
&+\delta[\tau-(t+(x+x_{2}^{1})/v_{g})]\} \bigg|^{2}\nonumber\\
=&\frac{\gamma\pi}{v_{g}^{2}}\bigg|c_{1}[t-(x+x_{1}^{1})/v_{g}]e^{-i\omega_{0}[t-(x+x_{1}^{1})/v_{g}]}\{\Theta[t-(t-(x+x_{1}^{1})/v_{g})]-\Theta[-(t-(x+x_{1}^{1})/v_{g})]\}\nonumber\\
&+c_{1}[t-(x-x_{1}^{1})/v_{g}]e^{-i\omega_{0}[t-(x-x_{1}^{1})/v_{g}]}\{\Theta[t-(t-(x-x_{1}^{1})/v_{g})]-\Theta[-(t-(x-x_{1}^{1})/v_{g})]\}\nonumber\\
&+c_{1}[t+(x-x_{1}^{1})/v_{g}]e^{-i\omega_{0}[t+(x-x_{1}^{1})/v_{g}]}\{\Theta[t-(t+(x-x_{1}^{1})/v_{g})]-\Theta[-(t+(x-x_{1}^{1})/v_{g})]\}\nonumber\\
&+c_{1}[t+(x+x_{1}^{1})/v_{g}]e^{-i\omega_{0}[t+(x+x_{1}^{1})/v_{g}]}\{\Theta[t-(t+(x+x_{1}^{1})/v_{g})]-\Theta[-(t+(x+x_{1}^{1})/v_{g})]\}\nonumber\\
&+c_{1}[t-(x+x_{2}^{1})/v_{g}]e^{-i\omega_{0}[t-(x+x_{2}^{1})/v_{g}]}\{\Theta[t-(t-(x+x_{2}^{1})/v_{g})]-\Theta[-(t-(x+x_{2}^{1})/v_{g})]\}\nonumber\\
&+c_{1}[t-(x-x_{2}^{1})/v_{g}]e^{-i\omega_{0}[t-(x-x_{2}^{1})/v_{g}]}\{\Theta[t-(t-(x-x_{2}^{1})/v_{g})]-\Theta[-(t-(x-x_{2}^{1})/v_{g})]\}\nonumber\\
&+c_{1}[t+(x-x_{2}^{1})/v_{g}]e^{-i\omega_{0}[t+(x-x_{2}^{1})/v_{g}]}\{\Theta[t-(t+(x-x_{2}^{1})/v_{g})]-\Theta[-(t+(x-x_{2}^{1})/v_{g})]\}\nonumber\\
&+c_{1}[t+(x+x_{2}^{1})/v_{g}]e^{-i\omega_{0}[t+(x+x_{2}^{1})/v_{g}]}\{\Theta[t-(t+(x+x_{2}^{1})/v_{g})]-\Theta[-(t+(x+x_{2}^{1})/v_{g})]\}\bigg|^{2}.
\end{align}
By introducing
\begin{align}
&R_{ij}= e^{-i\omega_{0}(t-t_{ij})}c_{1}(t-t_{ij})[\Theta(t-t_{ij})-\Theta(-t_{ij})], \label{S69}\\
&L_{ij}= e^{-i\omega_{0}(t+t_{ij})}c_{1}(t+t_{ij})[\Theta(t+t_{ij})-\Theta(t_{ij})], \label{S70}
\end{align}
where $t_{ij}\equiv [x+(-1)^{j+1}x_{i}^{1}]/v_{g}$ with indexes $i,j=1,2$, we can rewrite Eq.(\ref{S68}) to a more compact form
\begin{align}\label{S71}
I_{+}(x,t)=\frac{\gamma \pi}{v_{g}^{2}}|\underset{i,j}{\sum}(L_{ij}+R_{ij})|^{2},
\end{align}
as shown in Eq.(9) in the main text.
\subsection{Field intensity distribution for an anti-symmetric state}
For the case of anti-symmetric state $|-\rangle$, we would apply $\varphi_{R}(\omega,t)=-\varphi_{L}(\omega,t),c_{1}(t)=-c_{2}(t)$ to recalculate Eq.(\ref{S65}) and the field intensity distribution in this case reads
\begin{align}\label{S72}
I_{-}(x,t)=&\frac{4}{v_{g}^{2}}\int_{0}^{+\infty}d\omega_{1}\int_{0}^{+\infty}d\omega_{2}\,\varphi_{R}^{*}(\omega_{1},t)
\varphi_{R}(\omega_{2},t)\sin(\omega_{1}x/v_{g})\sin(\omega_{2}x/v_{g})e^{-i(\omega_{1}-\omega_{2})t}\nonumber\\
=&\frac{4}{v_{g}^{2}}\left|\int_{0}^{+\infty}d\omega \varphi_{R}(\omega,t)\sin(\omega x/v_{g})e^{-i\omega t}\right|^{2}.
\end{align}
Substituting the field excitation amplitude $\varphi_{R}(\omega,t)$ in Eq.(\ref{S3}) into Eq.(\ref{S72}), we obtain
\begin{align}\label{S73}
I_{-}(x,t)=&\frac{\gamma}{\pi v_{g}^{2}}\left|\int_{0}^{+\infty}d\omega\int_{0}^{t}d\tau\sum_{m,n=1,2}c_{m}(\tau)e^{-i\omega x_{n}^{m}/v_{g}}e^{i(\omega-\omega_{0})\tau}\sin(\omega x/v_{g})e^{-i\omega t}\right|^{2}\nonumber\\
=&\frac{\gamma}{\pi v_{g}^{2}}\left|\int_{0}^{+\infty}d\omega\int_{0}^{t}d\tau \sin(\omega x/v_{g})e^{-i\omega t}(e^{-i\omega x_{1}^{1}/v_{g}}+e^{-i\omega x_{2}^{1}/v_{g}}-e^{-i\omega x_{1}^{2}/v_{g}}-e^{-i\omega x_{2}^{2}/v_{g}})c_{1}(\tau)e^{i(\omega-\omega_{0})\tau}\right|^{2}.
\end{align}
For simplicity, we choose the center of the two braided giant atoms as the origin of coordinates such that $x_{1}^{1}=-x_{2}^{2},x_{2}^{1}=-x_{1}^{2}$, thus Eq.(\ref{S73}) becomes
\begin{align}\label{S74}
I_{-}(x,t)=\frac{4\gamma}{\pi v_{g}^{2}}\left|\int_{0}^{+\infty}d\omega\int_{0}^{t}d\tau \sin(\omega x/v_{g})e^{-i\omega t}[\sin(\omega x_{1}^{1}/v_{g})+\sin(\omega x_{2}^{1}/v_{g})]c_{1}(\tau)e^{i(\omega-\omega_{0})\tau}\right|^{2}.
\end{align}
After some algebra, we obtain
\begin{align}\label{S75}
I_{-}(x,t)=&\frac{\gamma\pi}{v_{g}^{2}}\bigg|\int_{0}^{t}d\tau c_{1}(\tau)e^{-i\omega_{0}\tau}\{\delta[\tau-(t-(x+x_{1}^{1})/v_{g})]-\delta[\tau-(t-(x-x_{1}^{1})/v_{g})]-\delta[\tau-(t+(x-x_{1}^{1})/v_{g})]\nonumber\\
&+\delta[\tau-(t+(x+x_{1}^{1})/v_{g})]+\delta[\tau-(t-(x+x_{2}^{1})/v_{g})]-\delta[\tau-(t-(x-x_{2}^{1})/v_{g})]-\delta[\tau-(t+(x-x_{2}^{1})/v_{g})]\nonumber\\
&+\delta[\tau-(t+(x+x_{2}^{1})/v_{g})]\}\bigg|^{2}\nonumber\\
=&\frac{\gamma\pi}{v_{g}^{2}}\bigg|c_{1}[t-(x+x_{1}^{1})/v_{g}]e^{-i\omega_{0}[t-(x+x_{1}^{1})/v_{g}]}\{\Theta[t-(t-(x+x_{1}^{1})/v_{g})]-\Theta[-(t-(x+x_{1}^{1})/v_{g})]\}\nonumber\\
&-c_{1}[t-(x-x_{1}^{1})/v_{g}]e^{-i\omega_{0}[t-(x-x_{1}^{1})/v_{g}]}\{\Theta[t-(t-(x-x_{1}^{1})/v_{g})]-\Theta[-(t-(x-x_{1}^{1})/v_{g})]\}\nonumber\\
&-c_{1}[t+(x-x_{1}^{1})/v_{g}]e^{-i\omega_{0}[t+(x-x_{1}^{1})/v_{g}]}\{\Theta[t-(t+(x-x_{1}^{1})/v_{g})]-\Theta[-(t+(x-x_{1}^{1})/v_{g})]\}\nonumber\\
&+c_{1}[t+(x+x_{1}^{1})/v_{g}]e^{-i\omega_{0}[t+(x+x_{1}^{1})/v_{g}]}\{\Theta[t-(t+(x+x_{1}^{1})/v_{g})]-\Theta[-(t+(x+x_{1}^{1})/v_{g})]\}\nonumber\\
&+c_{1}[t-(x+x_{2}^{1})/v_{g}]e^{-i\omega_{0}[t-(x+x_{2}^{1})/v_{g}]}\{\Theta[t-(t-(x+x_{2}^{1})/v_{g})]-\Theta[-(t-(x+x_{2}^{1})/v_{g})]\}\nonumber\\
&-c_{1}[t-(x-x_{2}^{1})/v_{g}]e^{-i\omega_{0}[t-(x-x_{2}^{1})/v_{g}]}\{\Theta[t-(t-(x-x_{2}^{1})/v_{g})]-\Theta[-(t-(x-x_{2}^{1})/v_{g})]\}\nonumber\\
&-c_{1}[t+(x-x_{2}^{1})/v_{g}]e^{-i\omega_{0}[t+(x-x_{2}^{1})/v_{g}]}\{\Theta[t-(t+(x-x_{2}^{1})/v_{g})]-\Theta[-(t+(x-x_{2}^{1})/v_{g})]\}\nonumber\\
&+c_{1}[t+(x+x_{2}^{1})/v_{g}]e^{-i\omega_{0}[t+(x+x_{2}^{1})/v_{g}]}\{\Theta[t-(t+(x+x_{2}^{1})/v_{g})]-\Theta[-(t+(x+x_{2}^{1})/v_{g})]\}\bigg|^{2}.
\end{align}
Applying the definitions in Eqs.(\ref{S69})-(\ref{S70}), then the field intensity distribution for anti-symmetric state of giant atoms is of the form
\begin{align}\label{S76}
I_{-}(x,t)=\frac{\gamma \pi}{v_{g}^{2}}|\underset{i,j}{\sum}(-1)^{j}(L_{ij}+R_{ij})|^{2}.
\end{align}
Importantly, each term in Eq.(\ref{S75}) has an intuitive physical picture as explained in detail in the main text.

\section{DYNAMICAL MANIPULATION OF OUTPUT FIELD}
We now investigate the dynamic regulation of output field due to the adjustability of superconducting atomic energy splitting. To show this process, we assume that there is a detector at $x_{d}=x_{2}^{2}+x_{0}$ $(x_{0}>0)$. Then we change the energy level splitting of atoms after a proper time, through which the standing wave boundary condition of sub-radiant states is broken. Here, we define the field annihilation operator $\hat{B}=\sqrt{\frac{2}{\pi}}\int dk\,\hat{b}_{k}e^{ikx}$ so that the field amplitude $\varphi(x,t)=\langle g,g,\{0\}|\hat{B}|\psi(t)\rangle$ reads
\begin{align}\label{S77}
\varphi(x,t)=&-ig_{k}\sqrt{\frac{2}{\pi}}\int dk\int_{0}^{t}dse^{ikx}e^{-iv_{g}(k-k_{0})(t-s)}[c_{1}(s)(e^{-ikx_{1}^{1}}+e^{-ikx_{2}^{1}})+c_{2}(s)(e^{-ikx_{1}^{2}}+e^{-ikx_{2}^{2}})]\nonumber\\
=&-ig_{k}\sqrt{\frac{2}{\pi}}\frac{2\pi}{v_{g}}\int_{0}^{t}ds\{c_{1}(s)e^{iv_{g}k_{0}(t-s)}[\delta(\frac{x-x_{1}^{1}}{v_{g}}-t+s)+\delta(\frac{x-x_{2}^{1}}{v_{g}}-t+s)]\nonumber\\
&+c_{2}(s)e^{iv_{g}k_{0}(t-s)}[\delta(\frac{x-x_{1}^{2}}{v_{g}}-t+s)+\delta(\frac{x-x_{2}^{2}}{v_{g}}-t+s)]\}.
\end{align}
We first consider the case of two separate giant atoms coupled to a 1D bosonic field, and the relevant position parameters are $x_{1}^{1}=-\frac{3}{2}d,x_{2}^{1}=-\frac{1}{2}d,x_{1}^{2}=\frac{1}{2}d,x_{2}^{2}=\frac{3}{2}d$. Then the field amplitude of the detector located at $\bar{x}=x_{0}+\frac{3}{2}d$ can be described by
\begin{align}\label{S78}
\varphi_{s}(\bar{x},\overline{t})=&-2i\sqrt{\frac{\gamma}{v_{g}}}e^{ik_{0}x_{0}}[e^{3i\varphi}c_{1}(\overline{t}-3\Delta t)\Theta(\overline{t}-3\Delta t)+e^{2i\varphi}c_{1}(\overline{t}-2\Delta t)\Theta(\overline{t}-2\Delta t)\nonumber\\
&+e^{i\varphi}c_{2}(\overline{t}-\Delta t)\Theta(\overline{t}-\Delta t)+c_{2}(\overline{t})\Theta(\overline{t})],
\end{align}
where we have defined $\overline{t}=t-x_{0}/v_{g}$ for simplicity. In the another case of two braided giant atoms, the field amplitude can be obtained by applying the similar procedures, i.e.,
\begin{align}\label{S79}
\varphi_{b}(\bar{x},\bar{t})=&-2i\sqrt{\frac{\gamma}{v_{g}}}e^{ik_{0}x_{0}}[e^{3i\varphi}c_{1}(\overline{t}-3\Delta t)\Theta(\overline{t}-3\Delta t)+e^{i\varphi}c_{1}(\overline{t}-\Delta t)\Theta(\overline{t}-\Delta t)\nonumber\\
&+e^{2i\varphi}c_{2}(\overline{t}-2\Delta t)\Theta(\overline{t}-2\Delta t)+c_{2}(\overline{t})\Theta(\overline{t})].
\end{align}
Or equivalently, we can rewrite Eqs.(\ref{S78})-(\ref{S79}) in a more compact way $\varphi(\bar{x},\overline{t})\rightarrow \varphi(\overline{t})$:
\begin{align}
&\varphi_{s}(\bar{t})=-\frac{2i}{\sqrt{\gamma v_{g}}}e^{ikx_{0}}[F_{1,3}(\overline{t})+F_{1,2}(\overline{t})+F_{2,1}(\overline{t})+F_{2,0}(\overline{t})],\label{S80}\\
&\varphi_{b}(\bar{t})=-\frac{2i}{\sqrt{\gamma v_{g}}}e^{ikx_{0}}[F_{1,3}(\overline{t})+F_{1,1}(\overline{t})+F_{2,2}(\overline{t})+F_{2,0}(\overline{t})],\label{S81}
\end{align}
where $\varphi_{s/b}$ denotes the detected output field amplitude of two separate/braided giant atoms.

\end{document}